%% file: jcapexample.tex
\documentclass[a4paper,11pt]{article}
\pdfoutput=1

\usepackage{jcappub}
\usepackage[T1]{fontenc}

\usepackage{booktabs}
\usepackage{longtable}
\usepackage{float} \makeatletter
\newcommand{\continuedcaption}[1]{\@makecaption{\fnum@figure}{#1}}
\makeatother
\usepackage{pdflscape}
\usepackage{array}
\usepackage{multirow}
\usepackage{makecell}
\usepackage{etoolbox}
\usepackage{url}

\newcommand{\apj}{Astrophys. J.}
\newcommand{\apjl}{Astrophys. J. Lett.}
\newcommand{\apjs}{Astrophys. J. Suppl.}
\newcommand{\aap}{Astron. Astrophys.}
\newcommand{\mnras}{Mon. Not. Roy. Astron. Soc.}
\newcommand{\aj}{Astron. J.}
\newcommand{\physrep}{Phys. Rept.}

\newcommand{\jcap}{J. Cosmol. Astropart. Phys.}

\providecommand{\citep}[1]{\cite{#1}}
\providecommand{\citet}[1]{ref.~\cite{#1}}

\newbool{showextra}
\setbool{showextra}{false}
\makeatletter
\newcolumntype{H}{>{\setbox0=\hbox\bgroup}c<{\egroup}@{}}
\makeatother
\ifbool{showextra}{\newcolumntype{V}{c}}{\newcolumntype{V}{H}}

\newcommand{\specfigdir}{Results/2026-08-02_v06_spectral-final/figure_set}
\newcommand{\specfigwidth}{0.32\textwidth}
\newcommand{\specfig}[1]{\includegraphics[width=\specfigwidth]{\specfigdir/\csname specfile#1\endcsname.pdf}}
\expandafter\def\csname specfile080825C\endcsname{grb080825c_band_cutoff}
\expandafter\def\csname specfile080916C\endcsname{grb080916c_band_cutoff}
\expandafter\def\csname specfile081215A\endcsname{grb081215a_band_cutoff}
\expandafter\def\csname specfile090217A\endcsname{grb090217a_cpl_cpl}
\expandafter\def\csname specfile090227B\endcsname{grb090227b_cpl_cpl}
\expandafter\def\csname specfile090323A\endcsname{grb090323a_cpl_cpl}
\expandafter\def\csname specfile090328A\endcsname{grb090328a_band_cutoff}
\expandafter\def\csname specfile090510A\endcsname{grb090510a_band_cpl}
\expandafter\def\csname specfile090902B\endcsname{grb090902b_cpl_cpl}
\expandafter\def\csname specfile090926A\endcsname{grb090926a_band_cutoff}
\expandafter\def\csname specfile091031A\endcsname{grb091031a_band_cutoff}
\expandafter\def\csname specfile100116A\endcsname{grb100116a_band_cutoff}
\expandafter\def\csname specfile100225A\endcsname{grb100225a_band_cutoff}
\expandafter\def\csname specfile100724B\endcsname{grb100724b_band_cutoff}
\expandafter\def\csname specfile100826A\endcsname{grb100826a_band_cutoff}
\expandafter\def\csname specfile101123A\endcsname{grb101123a_band_cutoff}
\expandafter\def\csname specfile110328B\endcsname{grb110328b_band_cutoff}
\expandafter\def\csname specfile110529A\endcsname{grb110529a_cpl_cpl}
\expandafter\def\csname specfile110721A\endcsname{grb110721a_band_cutoff}
\expandafter\def\csname specfile110731A\endcsname{grb110731a_cpl_cpl}
\expandafter\def\csname specfile120226A\endcsname{grb120226a_band_cutoff}
\expandafter\def\csname specfile120328B\endcsname{grb120328b_band_cutoff}
\expandafter\def\csname specfile120709A\endcsname{grb120709a_cpl_cpl}
\expandafter\def\csname specfile121011A\endcsname{grb121011a_cplpiv}
\expandafter\def\csname specfile130305A\endcsname{grb130305a_band_cutoff}
\expandafter\def\csname specfile130310A\endcsname{grb130310a_cpl_cpl}
\expandafter\def\csname specfile130504C\endcsname{grb130504c_band_cutoff}
\expandafter\def\csname specfile130518A\endcsname{grb130518a_band_cutoff}
\expandafter\def\csname specfile130606B\endcsname{grb130606b_band_cutoff}
\expandafter\def\csname specfile130828A\endcsname{grb130828a_cpl_cpl}
\expandafter\def\csname specfile131108A\endcsname{grb131108a_cpl_cpl}
\expandafter\def\csname specfile131216A\endcsname{grb131216a_band_cutoff}
\expandafter\def\csname specfile131231A\endcsname{grb131231a_band_cutoff}
\expandafter\def\csname specfile140110A\endcsname{grb140110a_cpl_cpl}
\expandafter\def\csname specfile140206B\endcsname{grb140206b_band_cutoff}
\expandafter\def\csname specfile140619B\endcsname{grb140619b_cpl_cpl}
\expandafter\def\csname specfile141028A\endcsname{grb141028a_band_cutoff}
\expandafter\def\csname specfile141207A\endcsname{grb141207a_cpl_cpl}
\expandafter\def\csname specfile141222A\endcsname{grb141222a_cpl_cpl}
\expandafter\def\csname specfile150210A\endcsname{grb150210a_cpl_cpl}
\expandafter\def\csname specfile150416A\endcsname{grb150416a_cpl_cpl}
\expandafter\def\csname specfile150523A\endcsname{grb150523a_cpl_cpl}
\expandafter\def\csname specfile151006A\endcsname{grb151006a_cpl_cpl}
\expandafter\def\csname specfile160509A\endcsname{grb160509a_band_cutoff}
\expandafter\def\csname specfile160625B\endcsname{grb160625b_cpl_cpl}
\expandafter\def\csname specfile160709A\endcsname{grb160709a_cpl_cpl}
\expandafter\def\csname specfile160816A\endcsname{grb160816a_cpl_cpl}
\expandafter\def\csname specfile160821A\endcsname{grb160821a_band_cutoff}
\expandafter\def\csname specfile160905A\endcsname{grb160905a_cpl_cpl}
\expandafter\def\csname specfile160910A\endcsname{grb160910a_band_cutoff}
\expandafter\def\csname specfile170214A\endcsname{grb170214a_band_cutoff}
\expandafter\def\csname specfile170405A\endcsname{grb170405a_band_cutoff}
\expandafter\def\csname specfile170424A\endcsname{grb170424a_band_cutoff}
\expandafter\def\csname specfile180703A\endcsname{grb180703a_band_cutoff}
\expandafter\def\csname specfile180720B\endcsname{grb180720b_band_cutoff}
\expandafter\def\csname specfile190114C\endcsname{grb190114c_cpl_cpl}
\expandafter\def\csname specfile190122A\endcsname{grb190122a_cplpiv}
\expandafter\def\csname specfile190531B\endcsname{grb190531b_cpl_cpl}
\expandafter\def\csname specfile200101A\endcsname{grb200101a_band_cutoff}
\expandafter\def\csname specfile200313A\endcsname{grb200313a_band_cutoff}
\expandafter\def\csname specfile200412B\endcsname{grb200412b_band_cutoff}
\expandafter\def\csname specfile200829A\endcsname{grb200829a_band_cutoff}
\expandafter\def\csname specfile201104A\endcsname{grb201104a_cpl_cpl}
\expandafter\def\csname specfile210410A\endcsname{grb210410a_cpl_cpl}
\expandafter\def\csname specfile210619B\endcsname{grb210619b_band_cutoff}
\expandafter\def\csname specfile220101A\endcsname{grb220101a_band_cutoff}
\expandafter\def\csname specfile220107B\endcsname{grb220107b_cpl_cpl}
\expandafter\def\csname specfile220209A\endcsname{grb220209a_band_cutoff}
\expandafter\def\csname specfile220527A\endcsname{grb220527a_band_cpl}
\expandafter\def\csname specfile221023A\endcsname{grb221023a_cpl_cpl}
\expandafter\def\csname specfile221027A\endcsname{grb221027a_band_cutoff}
\expandafter\def\csname specfile221209A\endcsname{grb221209a_band_cutoff}
\expandafter\def\csname specfile230512A\endcsname{grb230512a_cpl_cpl}
\expandafter\def\csname specfile240118A\endcsname{grb240118a_band_cutoff}
\expandafter\def\csname specfile240825A\endcsname{grb240825a_band_cpl}
\expandafter\def\csname specfile240905B\endcsname{grb240905b_cpl_cpl}
\expandafter\def\csname specfile250716A\endcsname{grb250716a_band_cutoff}

\title{High-energy spectral cutoffs in the prompt emission of \textit{Fermi}
gamma-ray bursts: bulk Lorentz factors, emission radii, and a cutoff--peak energy relation}

\author[a]{Yuan-Yuan Zuo,}
\author[a,1]{and Yuan-Chuan Zou\note[1]{Corresponding author.}}

\affiliation[a]{School of Physics, Huazhong University of Science and Technology,\\
Wuhan 430074, China}

\emailAdd{d202480118@hust.edu.cn}
\emailAdd{zouyc@hust.edu.cn}

\abstract{The high-energy end of the gamma-ray burst (GRB) prompt spectrum carries information on the physical
conditions of the relativistic outflow, but the number of well-characterized spectral cutoffs is small.
We present a systematic search for high-energy spectral cutoffs in the time-integrated prompt spectra
of 139 GRBs observed by \textit{Fermi} between 2008 July and 2025 December, selected to have broadband
coverage either through a LAT detection or through bright BGO emission.
Joint GBM, LLE, and LAT spectra over $T_{90}$ are fitted with four empirical models and compared using
the Bayesian information criterion, yielding 106 bursts with well-constrained cutoff energies $E_c$.
An empirical two-component Gaussian mixture model applied to $\log_{10}(E_c/{\rm MeV})$, separates a low-$E_c$ group from
the main distribution at $E_c=2.12$~MeV; the 77 bursts lying firmly above this boundary are interpreted
within the internal $\gamma\gamma$ pair-opacity framework.
Combining $E_c$ with wavelet-based minimum variability timescales (median $0.119$~s), we infer bulk
Lorentz factors of $11.7$--$2.78\times10^{3}$ (median 145) and emission radii of
$\sim10^{13}$--$10^{15}~{\rm cm}$.
Neither the $\Gamma$--$L_{\rm iso}$ nor the $\Gamma$--$E_{\rm iso}$ relation is statistically
significant for the full sample, whereas the 27 bursts with measured redshifts show a moderate
$\Gamma$--$L_{\rm iso}$ correlation with a slope of $0.36_{-0.23}^{+0.22}$.
We further find a strong rest-frame correlation between the cutoff and peak energies,
$E_{c,z}\propto E_{p,z}^{2.15_{-0.97}^{+0.96}}$, for the 26 bursts with measured redshifts, which
persists after controlling for redshift.}


\begin{document}
\maketitle
\raggedbottom

\section{Introduction}
\label{sec:intro}

Gamma-ray burst (GRB) prompt emission is generally attributed to highly relativistic outflows, but the physical conditions and dissipation processes responsible for the observed radiation remain uncertain.
The high luminosities and rapid variability of GRBs imply compact emitting regions with high photon densities~\cite{1999PhR...314..575P}.
The observed nonthermal emission extending to high energies therefore requires relativistic expansion to reduce the internal opacity~\cite{1986ApJ...308L..47G,1986ApJ...308L..43P,2001ApJ...555..540L}.
Energy dissipation within the outflow may occur through internal shocks produced by collisions between relativistic shells or through magnetic reconnection in a magnetized outflow~\cite{1997ApJ...490...92K,2011ApJ...726...90Z}.
The bulk Lorentz factor $\Gamma$ and emission radius $R$ are two key quantities for describing the prompt-emission region.
The Lorentz factor affects the transformation of photon energies and the optical depth of the emitting region, while the emission radius determines where the dissipation occurs within the relativistic outflow~\cite{2001ApJ...555..540L}.
The location of the dissipation region relative to the photosphere affects the optical conditions and can modify the observed prompt spectrum~\cite{2005ApJ...628..847R,2006ApJ...642..995P}.

The broad energy coverage of the Gamma-ray Burst Monitor (GBM) and Large Area Telescope (LAT) on board \textit{Fermi} has shown that GRB prompt spectra have diverse properties above the MeV band~\cite{2013ApJS..209...11A,2019ApJ...878...52A}.
In some GRBs, the high-energy emission is consistent with an extension of the main prompt component, whereas in others an additional hard spectral component is required~\cite{2010ApJ...716.1178A,2011ApJ...729..114A,2025A&A...700A..88M}.
Several mechanisms have been proposed for the high-energy emission, including leptonic processes such as synchrotron and inverse-Compton radiation, hadronic processes, and synchrotron emission from the external forward shock~\cite{2009A&A...498..677B,2009ApJ...699..953A,2010MNRAS.409..226K}. 
The high-energy component can itself exhibit spectral steepening or a cutoff, which may reflect intrinsic spectral curvature, changes in the radiation process, or attenuation of high-energy photons within the emitting region~\cite{2009A&A...498..677B,2011ApJ...729..114A}.
In parallel, photons in this high-energy regime have also been used to probe Lorentz invariance violation through possible energy-dependent propagation effects \cite{2008JCAP...01..031J,2024JCAP...04..060Y}.

For the spectral cutoffs considered here, one possible origin is internal $\gamma\gamma$ attenuation, in which high-energy photons interact with lower-energy photons through $\gamma\gamma$ pair production.
The corresponding opacity depends strongly on the bulk Lorentz factor, and the escape of the highest-energy photons has long been used to constrain lower limits on $\Gamma$~\cite{1997ApJ...491..663B,2001ApJ...555..540L,2009Sci...323.1688A}.
When a high-energy cutoff is detected and interpreted as internal $\gamma\gamma$ attenuation, the cutoff energy can instead provide a constraint on $\Gamma$, and the variability timescale can then be used to estimate the emission radius~\cite{2011ApJ...729..114A,2015ApJ...806..194T}.
Internal pair attenuation therefore provides a physical basis for connecting the observed high-energy cutoff with the relativistic outflow.

High-energy spectral cutoffs have been reported in a few GRBs and have been used to constrain the bulk Lorentz factor of the emitting region~\cite{2015ApJ...806..194T,2018ApJ...864..163V,2023ApJ...943..145L,2024A&A...685A.166R}.
These studies are based on a small number of bursts, and the cutoff properties have not been characterized in a uniform way across a broadband sample.
A larger and uniformly analyzed sample is therefore needed, both to establish how the cutoff energy is distributed and to test whether the quantities inferred from it are related to other prompt-emission observables such as the spectral peak energy and the prompt energetics.

In this work, we perform a systematic study of high-energy spectral cutoffs in the time-integrated prompt spectra of a broadband-selected sample of 139 \textit{Fermi} GRBs and investigate their implications for relativistic outflows within the internal $\gamma\gamma$ pair-opacity framework where physically applicable.
The data reduction and spectral analysis are described in section~\ref{sec:data-analysis}, while section~\ref{sec:cutoff-analysis} presents the cutoff-energy distribution, the Lorentz-factor and emission-radius constraints derived from the pair-opacity interpretation, their relations with the prompt energetics, and the relation between the cutoff and spectral peak energies.
The physical implications of these results are discussed in section~\ref{sec:discussion}, and our conclusions are summarized in section~\ref{sec:conclusions}.
Throughout the paper we adopt the Planck18 cosmology~\cite{2020A&A...641A...6P}, quote spectral-parameter uncertainties at the 90\% confidence level, and use the notation $Q_n=Q/10^n$ in cgs units.

\section{Data analysis}\label{sec:data-analysis}
\subsection{Sample selection}

The GRBs analyzed in this work were observed by the \textit{Fermi} Gamma-ray Space Telescope, which carries the GBM and LAT.
\textit{Fermi}-GBM is a wide-field monitor consisting of twelve sodium iodide (NaI) detectors and two bismuth germanate (BGO) detectors. The NaI detectors are sensitive over $8$--$1000$~keV and are oriented in different directions, providing coverage of nearly the entire unocculted sky. The two BGO detectors are on opposite sides of the spacecraft and extend the energy coverage to $0.2$--$40$~MeV~\cite{2009ApJ...702..791M}. 
\textit{Fermi}-LAT is a pair-conversion telescope sensitive to gamma-ray photons from $20$~MeV to more than $300$~GeV~\cite{2009ApJ...697.1071A}. 
LAT Low-Energy (LLE) data can be used for GBM-triggered GRBs located within $90^\circ$ of the LAT boresight~\cite{2010arXiv1002.2617P}.

Since the goal of this work is to characterize the high-energy spectral steepening during the prompt emission, we primarily selected GRBs that were observed by both GBM and LAT from July 2008 to December 2025. 
To avoid excluding GRBs with strong MeV emission but without LAT detection, we also included BGO-bright events for which the background-subtracted peak count rate in the 0.85--1.5~MeV band of the 1~s binned BGO light curve exceeds 250~counts~s$^{-1}$.
This threshold is empirical and was chosen so that the BGO data alone provide sufficient signal above 1~MeV to constrain the high-energy spectral shape.
The final sample contains 139 GRBs.
GRB~221009A was not included because its extreme brightness caused severe instrumental effects in the GBM data, which can strongly affect spectral analysis~\cite{2023ApJ...952L..42L}.

All \textit{Fermi}-GBM and LLE data were obtained from the public \textit{Fermi} data archive \footnote{\url{https://heasarc.gsfc.nasa.gov/FTP/fermi/data}}. The LAT data were extracted and processed using the \texttt{GTBURST} \footnote{\url{https://fermi.gsfc.nasa.gov/ssc/data/analysis/scitools/gtburst.html}} software.
The trigger time and $T_{90}$ duration were obtained from the \textit{Fermi}-GBM Burst Catalog~\cite{2014ApJS..211...13V}\footnote{\url{https://heasarc.gsfc.nasa.gov/W3Browse/fermi/fermigbrst.html}}.
The GRB redshift information was compiled from GRBWeb, an IceCube-hosted GRB catalog\footnote{\url{https://user-web.icecube.wisc.edu/~grbweb_public/Summary_table/Summary_table.html}}.

\subsection{Data reduction}
\label{subsec:data reduction}
For each GRB in our sample, we performed a standard data reduction using \texttt{GTBURST}.
For the \textit{Fermi}-GBM spectral analysis, we selected two NaI detectors and one BGO detector with the smallest viewing angles, excluding detectors affected by potential solar-flare contamination.
The background intervals were chosen from time regions before and after the prompt emission episode and were fitted with polynomial functions to estimate the background contribution.
The source spectra, background spectra, and weighted response files were then generated for the subsequent spectral analysis. 
For the NaI detectors, we used the energy range from 10~keV to 900~keV, excluding the 30--40~keV interval to avoid the iodine K-edge at 33.17~keV. 
For the BGO detectors, we selected energy channels between 300~keV and 40~MeV.
The LLE data were reduced following the same procedure as the GBM data, using the 30~MeV--10~GeV energy range.

For the LAT data, we performed the standard unbinned likelihood analysis. 
LAT high-energy events in the 100~MeV--100~GeV energy range and within the $T_{90}$ interval were extracted from a $12^\circ$ region of interest (ROI) centered on the GRB position.
Because the zenith angle of the GRB position varies among different bursts during the prompt emission, we tested zenith-angle cuts of $90^\circ$, $100^\circ$, and $105^\circ$, and adopted the appropriate value for each burst to reduce contamination from the Earth's limb.
Events in the Transient class were selected using the \texttt{P8R3\_TRANSIENT020E\_V3} response function.
In the LAT likelihood analysis, the gamma-ray emission was modeled with a \texttt{PowerLaw2} spectral model, while the particle background and Galactic diffuse emission were described by \texttt{isotr\_template} and \texttt{fixed norm} templates, respectively.
We included LAT spectra in the joint spectral fit only when the test statistic (TS) satisfied $\mathrm{TS}\geq10$, which we adopt as an approximate high-energy detection criterion; for a single degree of freedom this corresponds to a significance of about $3\sigma$.
When this criterion was not met, the joint fit was performed without the LAT spectrum, and the effect of this choice on the results is examined in section~\ref{sec:discussion}.

\subsection{Minimum variability timescale}

GRB prompt emission light curves exhibit rapid variability over a broad range of timescales, which may reflect the activity of the central engine and the internal dissipation processes within the relativistic outflow~\cite{1997ApJ...490...92K}.
The minimum variability timescale (MVT) characterizes the shortest timescale of statistically significant variability in the light curve.
Wavelet analysis provides a multi-resolution description of complex light curves and can identify the transition between the intrinsic scaling region and the noise-dominated region~\cite{2000ApJ...537..264W, 2012MNRAS.425L..32M, 2013MNRAS.432..857M}.

Here we estimate the MVT, denoted as $\delta t$, for each GRB using Fermi-GBM time-tagged event (TTE) data and following the wavelet-based method of MacLachlan et al.~\cite{2013MNRAS.432..857M}.
For each burst, we extracted light curves from the three NaI detectors with the smallest viewing angles to the GRB position in the 8--1000~keV energy range and rebinned them with a time resolution of 200~$\mu$s.
We used a preburst segment to estimate the background variability and subtracted its wavelet variance from that of the burst segment.
For long GRBs with $T_{90}$ duration greater than 2~s, the preburst interval was defined as $[-20,-5]$~s relative to the trigger time, and the signal interval was taken from $-5$~s to $T_{90,\rm stop}+50$~s.
For short GRBs, the corresponding intervals were $[-15,-1]$~s and $[-1,T_{90,\rm stop}+10]$~s, respectively.
Before the wavelet decomposition, each light curve segment was cropped to the largest available dyadic length required for the discrete wavelet transform.
In this dyadic scale grid, each octave $j$ corresponds to a characteristic timescale of $T_{\rm bin}2^j$, where $T_{\rm bin}$ is the light-curve resolution.
The logscale diagrams of the signal and preburst segments were compared only over the octaves common to both segments.
The finest octave and the largest common octave were excluded from the fits to reduce initialization and boundary effects. 
Reverse-tail concatenation was applied to reduce boundary discontinuities in the discrete wavelet transform~\cite{2013MNRAS.432..857M}.

We then applied a Haar wavelet transform~\cite{2013MNRAS.432..857M, 2016iwt..book.....A} to the light curve and constructed a logscale diagram from the scale-dependent variance of the wavelet detail coefficients.
In this diagram, the smallest timescales are generally dominated by white-noise fluctuations, while the intrinsic burst variability appears as a scaling region at larger timescales.
The MVT was determined from the intersection between the fitted noise-dominated region and the fitted scaling region~\cite{2013MNRAS.432..857M}.
If this intersection occurs at octave $j_{\rm int}$, the observer-frame MVT is given by $\delta t = T_{\rm bin}2^{j_{\rm int}}$.
The uncertainty of $\delta t$ was estimated through bootstrap realizations of the light curve.
In each realization, Poisson count perturbations and circular shifts were applied to account for statistical fluctuations and to reduce artifacts caused by the alignment of the light curve with the wavelet basis.

\subsection{Spectral fitting}
To characterize the high-energy spectral steepening, we fitted each GRB with four empirical spectral models: CPL$_{\rm piv}$, Band+cutoff, Band+CPL$_{\rm piv}$, and CPL+CPL$_{\rm piv}$. 
Here CPL denotes a cutoff power law, and the subscript ``piv'' indicates normalization at a pivot energy.
The functional forms used to construct these models are given in eqs.~\eqref{eq:cpl}--\eqref{eq:bandcut}:
\begin{equation}
N_{\rm CPL}(E)
=
N_0
E^{\alpha}
\exp\left(-\frac{E}{E_0}\right),
\label{eq:cpl}
\end{equation}
where $N_0$ is the normalization at 1~keV, $\alpha$ is the power-law photon index, and $E_0$ is the e-folding energy.

\begin{equation}
N_{\rm CPL_{\rm piv}}(E)
=
N_1
\left(\frac{E}{E_{\rm piv}}\right)^{\lambda}
\exp\left(-\frac{E}{E_c}\right),
\label{eq:cpl_piv}
\end{equation}
where $N_1$ is the normalization at the pivot energy, $E_{\rm piv}=1~{\rm MeV}$, $\lambda$ is the power-law photon index, and $E_c$ is the e-folding cutoff energy.

\begingroup
\begin{equation}
N_{\rm Band}(E)
=
\begin{cases}
N_0\left(\dfrac{E}{E_{\rm piv}}\right)^{\alpha}
\exp\left(-\dfrac{E}{E_0}\right),
&
E < E_{\rm a}, \\[8pt]
N_0\left(\dfrac{E_{\rm a}}{E_{\rm piv}}\right)^{\alpha-\beta}
\exp(\beta-\alpha)
\left(\dfrac{E}{E_{\rm piv}}\right)^{\beta},
&
E \geq E_{\rm a}, \\[8pt]
\end{cases}
\label{eq:band}
\end{equation}
\endgroup
where $N_0$ is the normalization constant, $E_{\rm piv}$ = 100~keV, $\alpha$ and $\beta$ are the low- and high-energy photon indices, respectively, $E_0$ is the characteristic energy, and $E_{\rm a}=(\alpha-\beta)E_0$.
Eq.~\eqref{eq:band} is the standard Band function~\cite{1993ApJ...413..281B}, and eq.~\eqref{eq:bandcut} replaces its high-energy power law by an exponentially cutoff power law.

\begingroup
\begin{equation}
N_{\rm Band+cutoff}(E)
=
\begin{cases}
N_0 E^{\alpha}
\exp\left(-\dfrac{E}{E_0}\right),
&
E < E_{\rm b}, \\[8pt]
N_0 E_{\rm b}^{\alpha-\beta}
\exp(\beta-\alpha)
E^{\beta}
\exp\left(-\dfrac{E}{E_c}\right),
&
E \geq E_{\rm b}, \\[8pt]
\end{cases}
\label{eq:bandcut}
\end{equation}
\endgroup
where $N_0$ is the normalization constant,
$E_0$ is the characteristic energy in keV,
$E_c$ is the e-folding energy, $\alpha$ and $\beta$ are the low- and high-energy photon indices, and $E_{\rm b} = {E_0E_c}(\alpha-\beta)/{(E_c-E_0)}$.

We performed time-integrated spectral fits for all 139 GRBs in the selected sample over their $T_{90}$ intervals using \texttt{XSPEC}\footnote{\url{https://heasarc.gsfc.nasa.gov/docs/software/xspec/}}.
For each burst, all available GBM, LLE, and LAT spectra were fitted jointly as separate data groups with their background and response files.
We adopted the Poisson data with Gaussian background statistic,
\texttt{pgstat}, for the GBM NaI and BGO spectra, and the Cash statistic,
\texttt{cstat}, for the LLE and LAT spectra.
The fitting energy ranges are given in section~\ref{subsec:data reduction}.
Each burst was fitted with the four spectral models described above.
To reduce the dependence of the fit on local minima, we repeated each fit using multiple sets of initial parameter values.
We then inspected the convergence status and parameter constraints of each model.
Parameter uncertainties were estimated at the 90\% confidence level using the \texttt{error} command in \texttt{XSPEC}.
Fits with parameters pegged at boundaries or with unconstrained confidence intervals were excluded from further consideration.
For bursts with at least one acceptable fit, the candidate models were compared using the Bayesian information criterion (BIC), calculated from the total fit statistic, the number of free parameters, and the number of spectral bins included in the joint fit.
We adopted the acceptable model with the lowest BIC as the preferred model.
After applying these fit-quality criteria and comparing models, 106 of the 139 GRBs yielded acceptable fits and were retained for the cutoff-energy analysis.

Because \texttt{pgstat} and \texttt{cstat} are likelihood-based statistics, their values cannot be interpreted in the same way as $\chi^2$, and the ratios of the total statistic to the number of degrees of freedom listed in tables~\ref{spectra} and \ref{tab:ec-nonhigh-properties} should therefore not be interpreted as reduced $\chi^2$.
A few of the brightest bursts nevertheless show unusually large total fit statistics; the possible origins and implications of these cases are discussed in section~\ref{sec:discussion}.

For the energetics calculation, we used the \texttt{cflux} convolution model in \texttt{XSPEC} to measure the energy flux of the full spectrum in the 10--1000~keV band.
These fluxes were used to derive the time-averaged isotropic-equivalent
luminosity $L_{\rm iso}$ and isotropic-equivalent energy
$E_{\rm iso}$ over the $T_{90}$ interval.

\section{High-energy cutoff analysis}\label{sec:cutoff-analysis}
A high-energy cutoff or spectral steepening in the prompt spectrum can arise from internal $\gamma\gamma$ absorption, an intrinsic cutoff in the radiating particle distribution, or other radiative effects~\cite{2009A&A...498..677B,2018ApJ...864..163V}.
In this work, we focus on the internal  $\gamma\gamma$ pair-production interpretation for GRBs with well-constrained cutoff energies.
In this scenario, high-energy photons are attenuated through $\gamma\gamma\rightarrow e^{+}e^{-}$ by lower-energy photons emitted from the same region, and the cutoff energy can be associated with the critical photon energy at which the pair-production optical depth becomes unity, $\tau_{\gamma\gamma}(E_c)=1$~\cite{1991ApJ...373..277K,2009Sci...323.1688A}.
The observed cutoff energy, together with the photon spectrum and variability timescale, can then be used to estimate the bulk Lorentz factor $\Gamma$ of the emitting outflow. 
In contrast, when no cutoff is detected, requiring the highest-energy observed photon to escape $\gamma\gamma$ absorption generally yields only a lower limit on $\Gamma$~\cite{1997ApJ...491..663B,2001ApJ...555..540L}.

\subsection{Pair-production opacity method}

Under the internal pair-production opacity interpretation, the optical depth for a high-energy photon observed at $E_c$ can be written as
\begingroup
\begin{equation}
\begin{split}
\tau_{\gamma\gamma}(E_c)
=&\,
\sigma_{\rm T}
\left(\frac{d_L}{R}\right)^2
\frac{E_{\rm ref} f(E_{\rm ref})}{(1+z)^{2(\eta+2)}}
F(\eta)
\left(
\frac{\Gamma^2 m_e^2c^4}
{E_cE_{\rm ref}}
\right)^{\eta+1},
\end{split}
\label{eq:tau_gg}
\end{equation}
\endgroup
where $\sigma_{\rm T}$ is the Thomson cross section, $z$ is the GRB redshift, and $d_L$ is the luminosity distance calculated using the Planck18 cosmology implemented in the \texttt{astropy} package~\cite{2018AJ....156..123A, 2020A&A...641A...6P}. 
The emission radius is taken as $R\simeq\Gamma^2c\delta t/(1+z)$, where $\delta t$ is the MVT measured from the burst light curve.
For GRBs without measured redshifts, we adopt $z=1$.
The time-integrated unattenuated high-energy photon spectrum associated with the fitted cutoff is written as $f(E)=f(E_{\rm ref})(E/E_{\rm ref})^\eta$, where $E_{\rm ref}$ is the reference energy and $\eta$ is the photon index.
The normalization $f(E_{\rm ref})$ is obtained by multiplying the photon-flux density at $E_{\rm ref}$ by the $T_{90}$ duration $\Delta T$ and has units of $\mathrm{photons\,cm^{-2}\,keV^{-1}}$.
For the Band+cutoff model, this spectrum corresponds to the high-energy $\beta$ part of the Band function, whereas for the other three models it corresponds to the $\lambda$ power-law part of the relevant CPL$_{\rm piv}$ component.
The dimensionless function $F(\eta)$ collects the dependence of the angle- and energy-averaged pair-production cross section on the photon index, and we adopt the approximation
$F(\eta)\approx0.597(-\eta)^{-2.30}$ for $-2.9\leq\eta\leq-1.0$ \cite{2009Sci...323.1688A,2015ApJ...806..194T}.
The $\Gamma_\tau$ solution is evaluated only when the relevant fitted photon index lies within this range.
Setting $\tau_{\gamma\gamma}(E_c)=1$ gives the Lorentz factor
\begingroup
\begin{equation}
\begin{split}
\Gamma_{\tau}
&=
\Biggl[
\sigma_{\rm T}
\left(\frac{d_L}{c\delta t}\right)^2
E_{\rm ref} f(E_{\rm ref})
F(\eta)
(1+z)^{-2(\eta+1)}
\left(
\frac{E_cE_{\rm ref}}{m_e^2c^4}
\right)^{-(\eta+1)}
\Biggr]^{\frac{1}{2(1-\eta)}} .
\end{split}
\label{eq:gamma_tau}
\end{equation}
\endgroup
The model-dependent quantities in eq.~\eqref{eq:gamma_tau} are assigned as follows. For the Band+cutoff model, $E_{\rm ref}=1~{\rm keV}$, $\eta=\beta$, and $f(E_{\rm ref})=N_0E_{\rm b}^{\alpha-\beta}\exp(\beta-\alpha)\Delta T$. For the Band+CPL$_{\rm piv}$, CPL+CPL$_{\rm piv}$, and single CPL$_{\rm piv}$ models, $E_{\rm ref}=E_{\rm piv}=1~{\rm MeV}$, $\eta=\lambda$, and $f(E_{\rm ref})=N_1\Delta T$.

Eq.~\eqref{eq:tau_gg} is a one-zone description, in which the target photons accumulated over the whole $T_{90}$ interval are assumed to fill a single region of size $R=\Gamma^2c\delta t/(1+z)$.
This is the same approximation adopted in previous cutoff-based analyses~\cite{2015ApJ...806..194T,2023ApJ...943..145L}, and it makes the results directly comparable, but it necessarily overestimates the target-photon density when $\Delta T\gg\delta t$ and therefore biases $\Gamma_\tau$ upward.
Calculations that follow the time-, space-, and direction-dependent photon field of a multi-zone outflow give systematically lower Lorentz factors, by a factor of a few, than the one-zone estimate~\cite{2008ApJ...677...92G,2012MNRAS.421..525H}.
The values derived below should therefore be interpreted with this systematic offset.

Because the gamma-ray spectra of GRBs generally decline steeply with energy,
$\gamma\gamma$ absorption at the cutoff is expected to be dominated by
interactions between photons at $E_c$ and lower-energy target photons
\cite{2010ApJ...709..525L}.
The cutoff photon must therefore be the higher-energy member of the interacting pair in the comoving frame, requiring $E'_c=E_c(1+z)/\Gamma \gtrsim \sqrt{2}\,m_ec^2$, which imposes an upper limit on the Lorentz factor,
\begin{equation}
\Gamma_{\max}
=
\frac{(1+z)E_c}{\sqrt{2}\,m_ec^2}.
\label{eq:gamma_max}
\end{equation}
Combining this kinematic constraint with the optical depth solution, the Lorentz factor adopted in our analysis is
\begin{equation}
\Gamma
=
\min\left(\Gamma_{\tau},\Gamma_{\max}\right).
\label{eq:gamma_final}
\end{equation}
The characteristic observer-frame energy of the lower-energy target photon is
\begin{equation}
E_{\rm ann}
=
\frac{2\Gamma^2(m_ec^2)^2}
{(1+z)^2E_c}.
\label{eq:eann}
\end{equation}
For the $\Gamma_\tau$-limited GRBs, $E_{\rm ann}$ characterizes the target-photon energy relevant to the opacity calculation, whereas for the $\Gamma_{\max}$-limited GRBs the kinematic limit gives $E_{\rm ann}=E_c$.

The two branches of eq.~\eqref{eq:gamma_final} have different physical interpretations.
When $\Gamma_\tau<\Gamma_{\max}$, the observed cutoff can be reproduced self-consistently by setting $\tau_{\gamma\gamma}(E_c)=1$, and $\Gamma_\tau$ is an estimate of the bulk Lorentz factor that depends on the photon spectrum, the variability timescale, and the distance.
When instead $\Gamma_\tau>\Gamma_{\max}$, no Lorentz factor satisfies both conditions: at any kinematically allowed $\Gamma$ the power-law estimate of the opacity at $E_c$ exceeds unity.
In this regime the attenuation is dominated by photons close to the pair-production threshold, where the power-law form of eq.~\eqref{eq:tau_gg} is no longer accurate and the true opacity is strongly suppressed, so the cutoff is expected to occur near $\Gamma\simeq\Gamma_{\max}$.
We therefore adopt $\Gamma_{\max}$ for these bursts, but we emphasize that it is a threshold estimate that depends only on $(1+z)E_c$ and is more properly regarded as an upper limit than as an independent measurement.
Because $R\propto\Gamma^2$, the same caveat applies to the emission radii of these bursts.

\subsection{Cutoff energy distribution}

\begin{figure}[tbp]
\centering
\includegraphics[width=0.56\textwidth]
{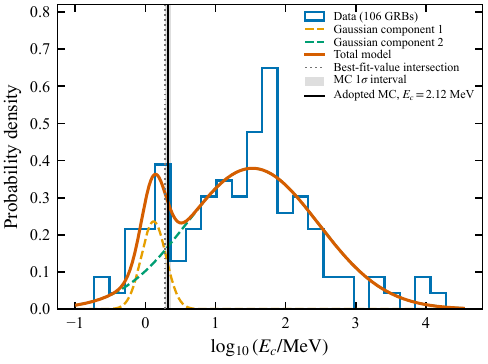}
\caption{
Gaussian mixture fit to the cutoff-energy distribution.
The dashed curves show the two weighted Gaussian components, and the orange solid curve shows their sum.
The dotted vertical line marks the weighted-density intersection obtained from the best-fit values, whereas the black line and gray interval mark the median and $1\sigma$ interval of the Monte Carlo boundary distribution.
}
\label{fig:ec-k2}
\end{figure}

The cutoff-energy distribution obtained from the spectral fits shows a possible low-$E_c$ concentration in addition to the main distribution, as illustrated in figure~\ref{fig:ec-k2}.
This low-energy tail includes three GRBs with best-fit cutoff energies below $0.511~{\rm MeV}$, two of which have measured redshifts and have $E_{c,z}<0.511~{\rm MeV}$.
These two cutoff features therefore do not satisfy the kinematic requirement of the internal $\gamma\gamma$ interpretation adopted in our analysis.
Although some of the other low-$E_c$ GRBs do not violate the pair-production threshold,
a relatively low cutoff energy generally favors the kinematic constraint and leads to smaller inferred Lorentz factors.
The radii are further reduced by their quadratic dependence on $\Gamma$, placing the inferred emission region in a very compact regime that may approach the photosphere.
For such low-energy cutoffs, identifying the observed steepening uniquely with internal $\gamma\gamma$ attenuation in a one-zone, optically thin emission region becomes less secure.
The observed steepening may instead reflect intrinsic spectral curvature or photospheric or subphotospheric emission~\cite{2011MNRAS.415.1663T,2015AdAst2015E...5P}.
Motivated by both the observed $E_c$ distribution and the concern over compact
emission regions,
we apply an empirical two-component Gaussian mixture model to $\log_{10}(E_c/{\rm MeV})$ and propagate the measurement uncertainties through Monte Carlo simulations.
Details of the classification are presented in appendix~\ref{app:ec-selection}.
Of the 106 GRBs with valid cutoff-energy measurements,
77 have a probability
$P_{\rm high}>0.99$
of lying above the adopted classification boundary $E_c = 2.12$~MeV.
These 77 GRBs are used in the subsequent pair-opacity analysis.
Such two-components separating at 2.12 MeV, indicates the two origins of the cutoffs, i.e., the lower ones are from radiation mechanism itself, while the higher ones are from the pair production.

\subsection{MVT and outflow properties}

\begin{figure}[tbp]
\centering
\includegraphics[width=0.86\textwidth]
{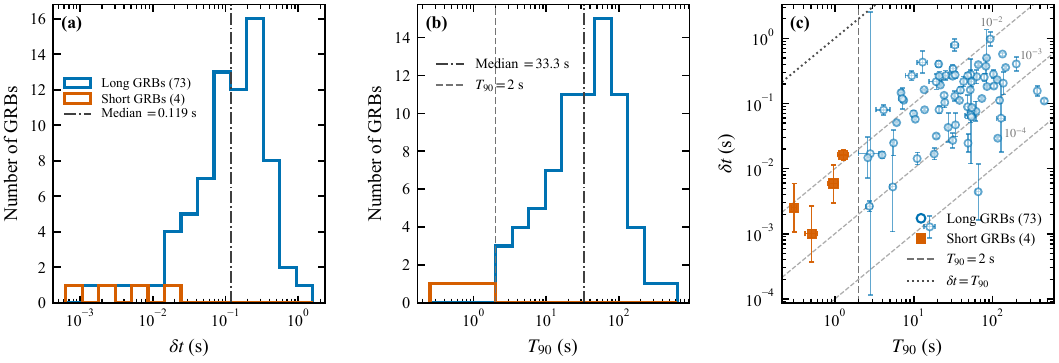}
\caption{Temporal distributions of the 77 GRBs used in the pair-opacity analysis.
Long and short GRBs are shown in blue and orange, respectively.
The dash-dotted lines in panels (a) and (b) mark the sample medians.
The dark dotted line in panel (c) marks $\delta t=T_{90}$, and the light dashed lines indicate $\delta t/T_{90}=10^{-2}$, $10^{-3}$, and $10^{-4}$.}
\label{fig:mvt-t90}
\end{figure}

For the 77 GRBs,
Table~\ref{tab:mvt-results} lists the adopted redshifts,
spectral-fitting intervals,
MVT measurements,
detector combinations used in the joint spectral fits,
and LAT TS for each burst.
The distributions of the MVT and $T_{90}$ values are shown in
figure~\ref{fig:mvt-t90}.
Among these,
73 are long and 4 are short GRBs.
The MVTs range from $1.0~{\rm ms}$ to $0.98~{\rm s}$,
with a median of $0.119~{\rm s}$.
For the 73 long GRBs,
the median is $0.130~{\rm s}$,
comparable to the $0.134~{\rm s}$ reported for long \textit{Fermi}-GBM GRBs
by Golkhou et al.~\cite{2015ApJ...811...93G},
although the samples and estimators differ.
The $T_{90}$ values range from $0.304$ to $453.4~{\rm s}$,
with a median of $33.3~{\rm s}$.
The MVT estimate is shorter than $T_{90}$ for every GRB,
and the median $\delta t/T_{90}$ is $4.1\times10^{-3}$.
Most GRBs in panel (c) of figure~\ref{fig:mvt-t90} lie between
$\delta t/T_{90}=10^{-3}$ and $10^{-2}$.

\begin{figure}[tbp]
\centering
\includegraphics[width=0.94\textwidth]{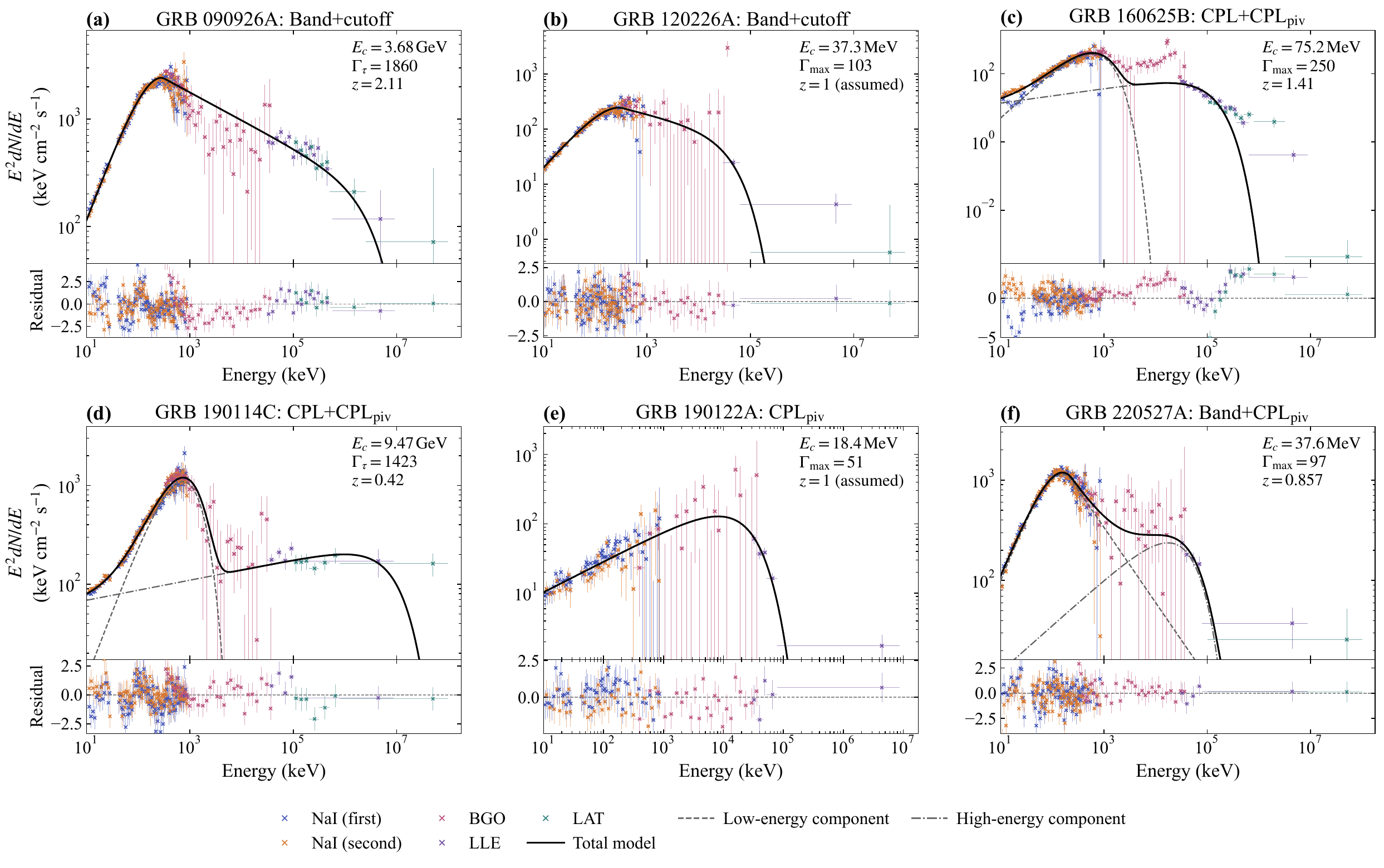}
\caption{Examples of the prompt spectral fits.
Colored crosses and error bars denote the individual detectors, as identified in the legend.
The smooth black solid curve represents the total model.
For the two-component fits, the gray dashed and dash-dotted curves show the low- and high-energy components, respectively.
The upper-right annotation in each panel gives the best-fit cutoff energy,
the Lorentz factor, and the adopted redshift.
The fits for the remaining 71 GRBs are shown in appendix~\ref{app:spectral-figures}.}
\label{fig:spectral-examples}
\end{figure}

The preferred spectral model, best-fit spectral parameters, and the derived
$\Gamma$, $E_{\rm ann}$, and $R$ for each burst are reported in
table~\ref{spectra}.
Figure~\ref{fig:spectral-examples} shows six representative prompt spectral
fits, GRBs~090926A, 120226A, 160625B, 190114C, 190122A, and 220527A, spanning all four spectral models and both limiting constraints.
The fits for the remaining 71 GRBs are collected in appendix~\ref{app:spectral-figures}.

\begin{figure}[tbp]
\centering
\includegraphics[width=0.86\textwidth]
{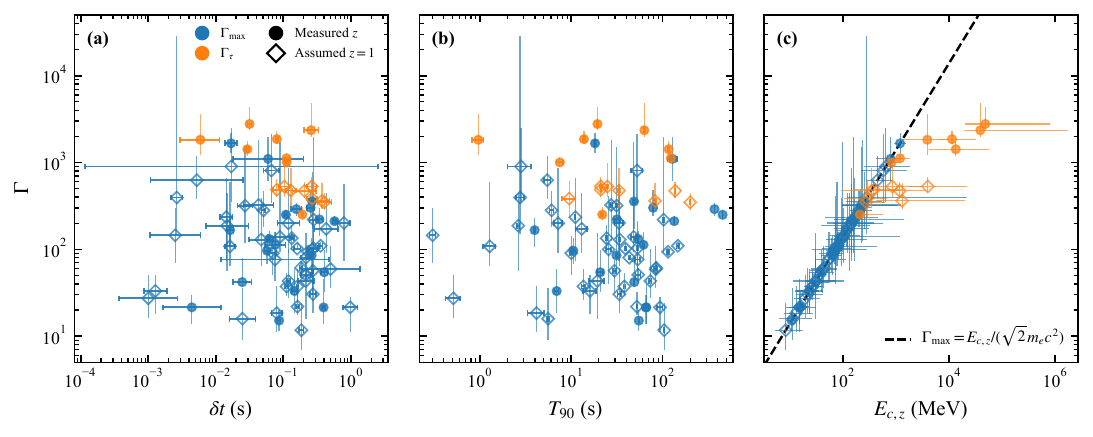}
\caption{Lorentz factor $\Gamma$ versus (a) $\delta t$, (b) $T_{90}$, and (c) $E_{c,z}$.
The colors distinguish the $\Gamma_{\max}$-limited and $\Gamma_\tau$-limited sources,
and filled circles and open diamonds denote measured redshifts and the
assumed $z=1$,
respectively.
The dashed line in panel (c) is eq.~\eqref{eq:gamma_max},
not a regression fit.}
\label{fig:gamma-time}
\end{figure}

The inferred Lorentz factors span $11.7$--$2.78\times10^3$, with a median of 145.
The final Lorentz factor constraint is set by $\Gamma_\tau$ for 16 GRBs and by $\Gamma_{\max}$ for 61, with median values of 530 and 103, respectively.
The two subsets should therefore be read differently: for the 16 $\Gamma_\tau$-limited bursts the Lorentz factor follows from the opacity condition and depends on the spectrum, the MVT, and the distance, whereas for the remaining 61 bursts it reduces to the kinematic scaling $\Gamma_{\max}=(1+z)E_c/(\sqrt{2}m_ec^2)$ and carries no independent information beyond $E_{c,z}$.
For comparison, Tang et al.~\cite{2015ApJ...806..194T} obtained $\Gamma\simeq52$--748 for eight GRBs from time-integrated cutoff spectra, while Li et al.~\cite{2023ApJ...943..145L} found $\Gamma\simeq60$--682 for 34 pulses in nine GRBs.
Our inferred values overlap the ranges reported in these studies, while the larger sample extends the $\Gamma$ distribution to both lower and higher values.
Figure~\ref{fig:gamma-time} compares $\Gamma$ with the MVT, $T_{90}$, and the rest-frame cutoff energy.
No clear trend is evident between $\Gamma$ and either temporal quantity in panels~(a) and (b).
This differs from the decline of the variability timescale with increasing $\Gamma$ reported by Sonbas et al.~\cite{2015ApJ...805...86S}. Their Lorentz-factor estimates were obtained independently of prompt spectral cutoffs, whereas $\Gamma$ is set by $\Gamma_{\max}\propto E_{c,z}$ for most of our sample, so the two relations are not directly comparable.
In panel (c), the dashed line marks the kinematic limit given by eq.~\eqref{eq:gamma_max}. 
Because $\Gamma_{\max}$ increases linearly with $E_{c,z}$ whereas $\Gamma_\tau$ depends much more weakly on the cutoff energy, the Lorentz-factor constraint is generally set by $\Gamma_{\max}$ at lower $E_{c,z}$ and is more likely to be set by $\Gamma_\tau$ at higher $E_{c,z}$.

\begin{figure}[tbp]
\centering
\includegraphics[width=0.78\textwidth]
{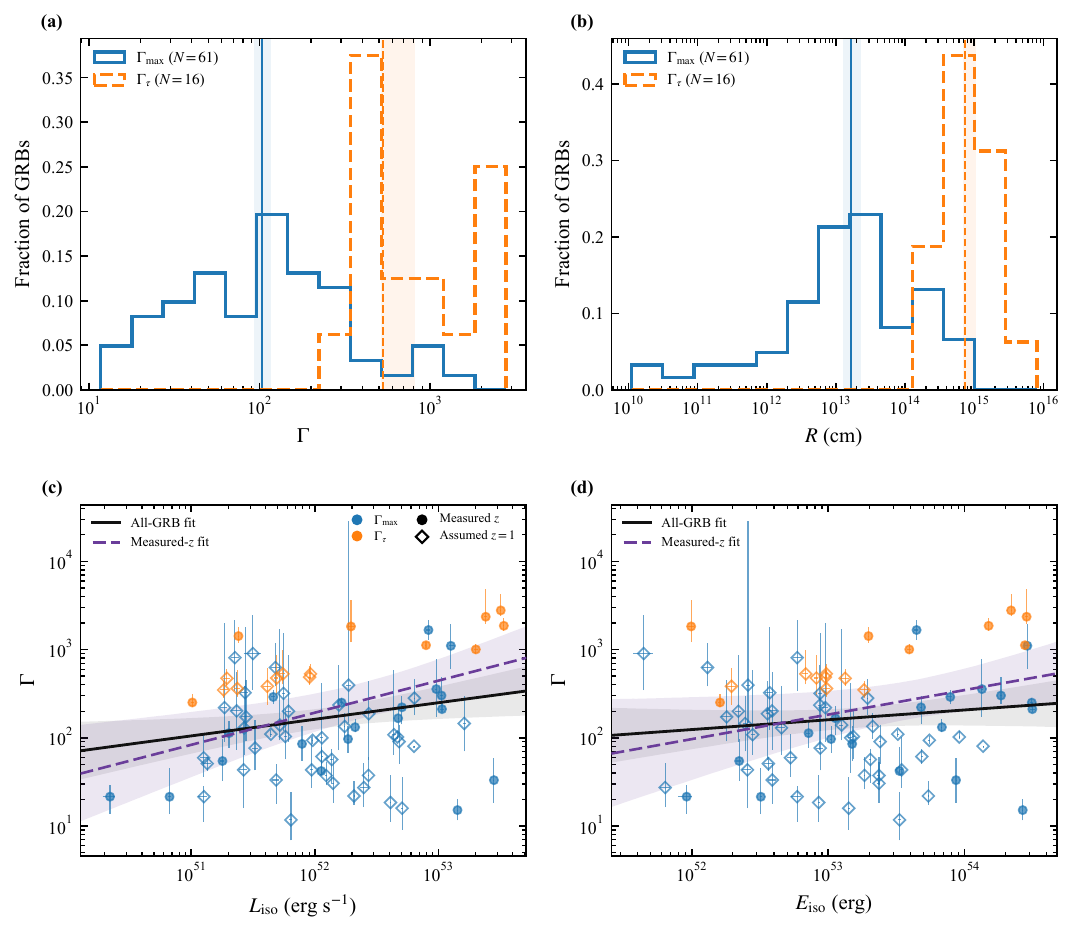}
\caption{Outflow and energetics results.
Panels (a) and (b) show the $\Gamma$ and $R$ distributions, normalized separately for the $\Gamma_{\max}$-limited and $\Gamma_\tau$-limited subsets.
The vertical lines mark the sample medians of the individual estimates.
Panels (c) and (d) show the $\Gamma$--$L_{\rm iso}$ and $\Gamma$--$E_{\rm iso}$ relations, respectively.
The black solid lines show the posterior-median regressions for all 77 GRBs, whereas the purple dashed lines show the corresponding fits for the 27 GRBs with measured redshifts.
The shaded regions show the 90\% credible intervals for the regression relations.
The Spearman coefficients are $\rho=0.13$ ($p=0.28$) and $\rho=0.07$ ($p=0.57$) for the full-sample $\Gamma$--$L_{\rm iso}$ and $\Gamma$--$E_{\rm iso}$ relations, respectively, and $\rho=0.47$ ($p=0.015$) and $\rho=0.35$ ($p=0.072$) for the corresponding relations in the measured-redshift subsample.
Colors and symbols are defined as in figure~\ref{fig:gamma-time}.}
\label{fig:outflow-energetics}
\end{figure}

Combined with the measured MVTs, the inferred Lorentz factors also constrain the characteristic emission radii of the prompt-emission region.
The upper panels of figure~\ref{fig:outflow-energetics} show the distributions of $\Gamma$ and $R$.
The $\Gamma_{\max}$-limited and $\Gamma_\tau$-limited GRBs have median radii of $1.67\times10^{13}$ and $7.35\times10^{14}~{\rm cm}$, respectively.
The regressions in figures~\ref{fig:outflow-energetics} and \ref{fig:ec-ep} are performed in logarithmic space, with the independent variable centered at its sample median, using a Bayesian linear-regression model that accounts for uncertainties in both variables and intrinsic scatter~\cite{2005physics...11182D}.
We report 90\% posterior intervals for the regression parameters together with Spearman coefficients and $p$-values, and repeat the fits for the measured-redshift subsamples.
For descriptive purposes, we follow Schober et al.~\cite{Schober2018} and refer to correlations with $|\rho|<0.10$, $0.10\leq|\rho|<0.40$, $0.40\leq|\rho|<0.70$, $0.70\leq|\rho|<0.90$, and $|\rho|\geq0.90$ as negligible, weak, moderate, strong, and very strong, respectively.
We adopt $p<0.05$ as the threshold for statistical significance.

For all 77 GRBs, the $\Gamma$--$L_{\rm iso}$ correlation is weak, while the $\Gamma$--$E_{\rm iso}$ correlation is negligible, with Spearman coefficients of $\rho=0.13$ ($p=0.28$) and $\rho=0.07$ ($p=0.57$), respectively.
Neither relation is statistically significant.
For the 27 GRBs with measured redshifts, the fitted relations become $\Gamma=10^{2.53_{-0.21}^{+0.21}}(L_{\rm iso,52}/4.72)^{0.36_{-0.23}^{+0.22}}$ and $\Gamma=10^{2.45_{-0.22}^{+0.22}}(E_{\rm iso,53}/4.79)^{0.28_{-0.27}^{+0.26}}$, with $\rho=0.47$ ($p=0.015$) and $\rho=0.35$ ($p=0.072$), respectively.
The measured-redshift subsample therefore shows a moderate and statistically significant $\Gamma$--$L_{\rm iso}$ correlation, whereas the $\Gamma$--$E_{\rm iso}$ correlation is still weak and does not reach statistical significance.
The exponents are consistent with the initial-Lorentz-factor relations $\Gamma_0$--$L_{\gamma,\rm iso}$ and $\Gamma_0$--$E_{\gamma,\rm iso}$ slopes of approximately 0.30 and 0.29 inferred independently from afterglow onset times by L\"{u} et al.~\cite{2012ApJ...751...49L}.
However, our measured-redshift slopes are shallower than the cutoff-based values of $0.52\pm0.13$ and $0.88\pm0.35$ reported for eight GRBs by Tang et al.~\cite{2015ApJ...806..194T}.
This difference may partly reflect the broader range of prompt-emission properties and the coexistence of $\Gamma_{\max}$- and $\Gamma_\tau$-limited GRBs, which may contribute to the scatter in the $\Gamma$--energetics relations.

\subsection{Cutoff--peak energy relation}
\label{subsec:ec-ep}

To explore whether the low-energy spectral peak and high-energy cutoff are related, we examine the $E_c$--$E_p$ relation for 75 GRBs, excluding the two single CPL$_{\rm piv}$ fits for which $E_p$ is not independent of $E_c$.
\begin{figure}[tbp]
\centering
\includegraphics[width=0.78\textwidth]
{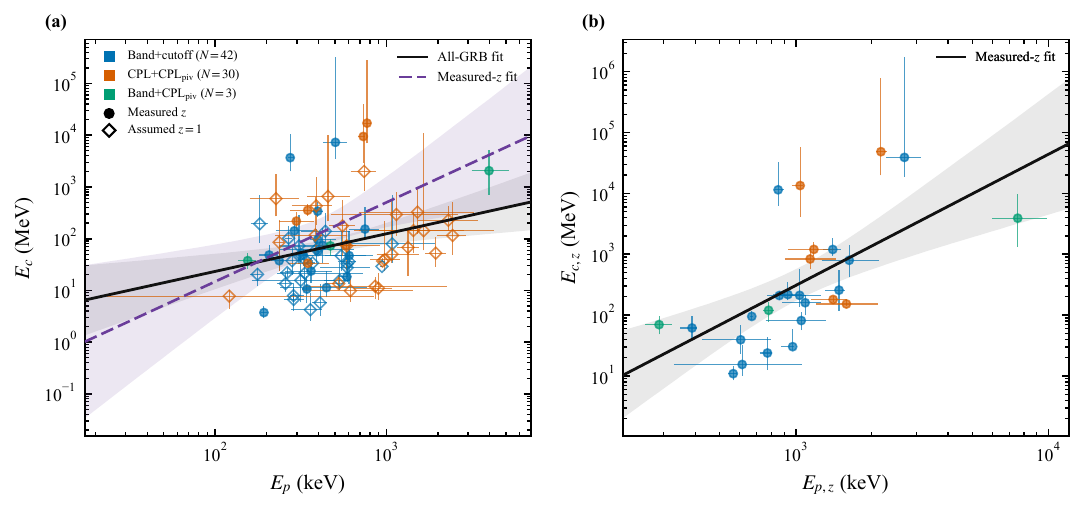}
\caption{Relation between the cutoff energy and the spectral peak energy.
Panel (a) shows the observer-frame $E_c$--$E_p$ distribution for 75 GRBs.
Colors identify the spectral models, while filled and open symbols distinguish measured and assumed redshifts, respectively.
The black solid line shows the posterior-median regression for all 75 GRBs, and the purple dashed line shows the fit for the 26 GRBs with measured redshifts, with Spearman coefficients of $\rho=0.25$ ($p=0.031$) and $\rho=0.38$ ($p=0.055$), respectively.
Panel (b) shows the rest-frame $E_{c,z}$--$E_{p,z}$ distribution and regression for the 26 GRBs with measured redshifts, with a slope of $2.15_{-0.97}^{+0.96}$ and a Spearman coefficient of $\rho=0.72$ ($p=8\times10^{-5}$).
The shaded regions denote the 90\% credible intervals of the regression relations.}
\label{fig:ec-ep}
\end{figure}
For all 75 GRBs, the observer-frame $E_c$--$E_p$ relation shows a weak but statistically significant positive correlation, with $\rho=0.25$ and $p=0.031$.
Restricting the sample to the 26 GRBs with measured redshifts, the correlation remains weak, with $\rho=0.38$ and $p=0.055$, and does not reach statistical significance.
However, after transforming to the rest frame, the same 26 GRBs show a strong and statistically significant positive correlation, $E_{c,z}/{\rm MeV}=10^{2.52_{-0.26}^{+0.26}}(E_{p,z}/1.04~{\rm MeV})^{2.15_{-0.97}^{+0.96}}$, with $\rho=0.72$ and $p=8\times10^{-5}$.
The partial Spearman correlation remains statistically significant after controlling for redshift, with $\rho=0.63$ and $p=7\times10^{-4}$, indicating that this correlation cannot be explained solely by the common redshift dependence of the two energies.
The fitted slope, $2.15_{-0.97}^{+0.96}$, is larger than unity, suggesting a relation steeper than simple proportionality between $E_{c,z}$ and $E_{p,z}$, although the uncertainty remains substantial.
Both plots are shown in Figure \ref{fig:ec-ep}.

Two systematic effects were examined before this relation is interpreted physically.
First, $E_p$ and $E_c$ are derived from the same joint spectral fit, so their local covariance could affect the inferred regression.
We account for this covariance using the Hessian matrix of the final spectral fits.
For the 26 GRBs with measured redshifts, setting the covariance to zero changes the median slope of the rest-frame relation only from 2.154 to 2.138, well within the uncertainty of the fitted slope.
Second, transforming to the rest frame multiplies both energies by the same factor $(1+z)$ and can strengthen a correlation.
The partial Spearman test controls for this common redshift dependence, and the correlation remains significant.

Within the internal $\gamma\gamma$-attenuation interpretation, $E_p$ and $E_c$ can be physically linked through the prompt photon field.
The cutoff energy is determined approximately by the condition $\tau_{\gamma\gamma}(E_c)\sim1$, for which the opacity depends on both the density and the spectral distribution of the target photons~\cite{1997ApJ...491..663B,2001ApJ...555..540L,2008MNRAS.384L..11G}.
For photons near $E_c$, the pair-production kinematics select a characteristic range of target-photon energies, determined mainly by $E_c$, $\Gamma$, and the collision geometry.
The photon density within this range is governed by the shape and normalization of the prompt spectrum, with $E_p$ setting the location of the spectral peak relative to the target-photon energy range.
Thus, $E_p$ influences the pair opacity through the target-photon distribution without directly determining either the target-photon energy or $E_c$.
Because $E_p$ and $E_c$ characterize different parts of the prompt spectrum, the persistence of the correlation after controlling for redshift, together with a fitted slope larger than unity, argues against a simple origin from the common redshift correction or a nearly constant $E_c/E_p$ ratio and supports a physical connection between the cutoff formation and the conditions shaping the prompt spectrum.

Previous studies of prompt spectral cutoffs have focused mainly on using $E_c$ to constrain the bulk Lorentz factor~\cite{2015ApJ...806..194T,2023ApJ...943..145L,2024A&A...685A.166R}.
To our knowledge, an $E_c$--$E_p$ correlation across a sample of GRBs has not been reported, although the small number of bursts with measured cutoffs in earlier work would have made such a relation difficult to detect.
Given the limited number of GRBs with measured redshifts and our use of time-integrated spectra, the generality and physical origin of this empirical relation should be further tested with larger redshift samples and time-resolved spectroscopy.

\section{Discussion} \label{sec:discussion}

In our analysis, the fitted cutoff is associated with different spectral components across the sample.
For 33 of the 77 GRBs, the preferred model contains a CPL$_{\rm piv}$ component in addition to the low-energy Band or CPL component, whereas for 42 GRBs the cutoff modifies the high-energy part of a single Band-like spectrum.
The remaining two GRBs are fitted by a single CPL$_{\rm piv}$ model.
This diversity is consistent with previous \textit{Fermi} studies showing that the emission above the MeV band may either extend the main prompt spectrum or appear as an additional hard component~\cite{2011ApJ...729..114A,2025A&A...700A..88M}.
Possible origins of an additional high-energy component include inverse-Compton radiation, hadronic processes, and emission from the early external shock~\cite{2009A&A...498..677B,2009ApJ...699..953A,2010MNRAS.409..226K}, although the phenomenological spectral decomposition alone does not determine its physical origin.

We interpret the high-energy cutoffs within the internal $\gamma\gamma$-attenuation framework and use them to constrain the bulk Lorentz factors and emission radii.
Several of these GRBs have been studied, allowing a direct comparison with earlier Lorentz-factor estimates.
For GRB~100724B and GRB~160509A, we obtain $\Gamma\simeq94$ and 290, respectively, broadly consistent with the $\Gamma\sim100$--400 range inferred for these bursts by Vianello et al.~\cite{2018ApJ...864..163V}.
For GRB~090926A and GRB~190114C, on the other hand, our estimates of $\Gamma\simeq1860$ and 1420 are higher than those inferred from shorter time intervals.
In GRB~090926A, the cutoff is most prominent during the narrow high-energy pulse, for which Ackermann et al.~\cite{2011ApJ...729..114A} inferred a Lorentz factor of a few hundred.
In GRB~190114C, the high-energy component shows evolving spectral attenuation during the early prompt phase~\cite{2020ApJ...890....9A}, and time-resolved analyses of the sub-GeV spectrum give Lorentz factors of a few hundred for the individual pulses~\cite{2020ApJ...903....9C}.
Beyond these well-studied examples, our analysis applies the same framework to all 77 GRBs, extending cutoff-based Lorentz-factor and emission-radius constraints to a substantially larger prompt-emission sample.
Because our spectra are integrated over the GBM $T_{90}$ interval, they combine multiple emission episodes whose spectral properties and physical conditions may evolve with time.
This is particularly relevant at high energies, where LAT emission commonly shows a delayed onset and can persist longer than the GBM emission~\cite{2013ApJS..209...11A,2019ApJ...878...52A}, so the time-integrated high-energy spectrum may include emission from different stages of the prompt evolution.
Time integration can also contribute to the large fit statistics of some bright bursts, since the goodness of fit of time-integrated \textit{Fermi}-GBM spectra has been found to worsen toward higher fluence when spectral evolution is not fully captured by empirical spectral models \cite{2012ApJS...199...19G}.
The resulting cutoff energies, Lorentz factors, and emission radii should therefore be regarded as effective quantities over the prompt interval.

In the full sample, neither the $\Gamma$--$L_{\rm iso}$ nor the $\Gamma$--$E_{\rm iso}$ relation is statistically significant.
For the GRBs with measured redshifts, the $\Gamma$--$L_{\rm iso}$ relation is moderate and statistically significant, whereas the $\Gamma$--$E_{\rm iso}$ relation remains weak and does not reach statistical significance.
The regression slopes for the GRBs with measured redshifts are broadly consistent with the $\Gamma_0$--$L_{\gamma,\rm iso}$ and $\Gamma_0$--$E_{\gamma,\rm iso}$ relations inferred from afterglow onset times by Liang et al.~\cite{2010ApJ...725.2209L} and L\"{u} et al.~\cite{2012ApJ...751...49L}, and with the larger compilation of Ghirlanda et al.~\cite{2018A&A...609A.112G}.
This agreement is noteworthy because the Lorentz factors are inferred independently from prompt spectral cutoffs in our analysis and from the afterglow onset in their study.
In contrast, our fitted slopes are shallower than the cutoff-based relations reported by Tang et al.~\cite{2015ApJ...806..194T}, which may reflect differences in sample selection and spectral analysis.
The lack of clear correlations in the full sample may also be affected by the assumed redshift for most GRBs, since both $L_{\rm iso}$ and $E_{\rm iso}$ depend strongly on the adopted distance.
The present results therefore suggest that the bulk Lorentz factor is not determined by the prompt energetics alone, but may also depend on the dynamical and radiative properties of the relativistic outflow.

Among the correlations examined here, the rest-frame $E_{c,z}$--$E_{p,z}$ relation stands out as a strong positive correlation.
This result is notable because $E_p$ and $E_c$ describe different parts of the prompt spectrum: $E_p$ marks the peak of the $\nu F_\nu$ spectrum, whereas $E_c$, under the internal $\gamma\gamma$ interpretation, marks the energy at which attenuation of high-energy photons becomes important.
The correlation remains significant after controlling for redshift, indicating that the common redshift dependence alone cannot account for the observed relation.
The result therefore suggests that the spectral peak and the physical conditions governing high-energy attenuation are not independent across GRBs.

The connection between $E_p$ and $E_c$ can be understood in the pair-opacity framework because $E_p$ affects the target photon field, while the cutoff also depends on the bulk Lorentz factor and emission radius.
In particular, larger values of $\Gamma$ or $R$ generally reduce the internal pair opacity, allowing higher-energy photons to escape.
The observed $E_{c,z}$--$E_{p,z}$ relation may therefore reflect the combined influence of the prompt photon field and the physical conditions of the relativistic outflow.
The best-fit slope above unity is consistent with these additional dependencies, although its substantial uncertainty does not allow a specific theoretical scaling to be established.
We therefore regard the measured relation as an empirical property of the present sample rather than a direct prediction of the pair-opacity model.
If confirmed with larger measured-redshift samples and time-resolved spectroscopy, this relation could provide a new observational link between the spectral peak and the physical conditions that set the high-energy cutoff.

The 29 GRBs excluded from the high-$E_c$ sample provide additional information on the possible origin of low-energy cutoff features.
As described in appendix~\ref{app:ec-selection}, 15 are classified as low-$E_c$ GRBs and 14 as boundary-crossing GRBs.
Two of the low-$E_c$ GRBs with measured redshifts have $E_{c,z}<0.511~{\rm MeV}$ and therefore cannot be consistently interpreted as internal $\gamma\gamma$ attenuation under the adopted kinematic requirement.
The boundary-crossing GRBs mainly reflect classification uncertainty near the empirical $E_c$ boundary.
Several other processes may produce high-energy spectral steepening in prompt GRB spectra.
Internal-shock calculations by Bo\v{s}njak et al.~\cite{2009A&A...498..677B} show that high-energy suppression can arise from internal $\gamma\gamma$ absorption or from Klein--Nishina effects in inverse-Compton emission, depending on the radiative conditions.
An intrinsic cutoff in the energy distribution of the emitting particles has also been considered as an alternative explanation for prompt spectral cutoffs~\cite{2018ApJ...864..163V}.
Radiative transfer in dissipative photospheres can produce strong spectral curvature and cutoff-like features through Comptonization~\cite{2016ApJ...831..175V}.
In addition, because our spectra are integrated over $T_{90}$, spectral evolution among different emission episodes can contribute to the observed time-integrated curvature, as demonstrated by time-resolved studies of individual GRBs~\cite{2018ApJ...864..163V,2023ApJ...943..145L}.
These alternatives are particularly relevant for the excluded low-$E_c$ GRBs and require time-resolved broadband spectroscopy to distinguish, since different mechanisms can produce similar cutoff-like features in time-integrated spectra but different spectral evolution.

A further systematic effect concerns the opacity calculation itself.
Eq.~\eqref{eq:tau_gg} treats the emission region as a single zone whose photon field is represented by the $T_{90}$-integrated spectrum and whose size follows from the MVT.
In a realistic outflow, high-energy photons propagate through a photon field produced at different radii and times, and detailed calculations show that the resulting Lorentz factors can be lower than the one-zone estimates by a factor of a few~\cite{2008ApJ...677...92G,2011ApJ...726L...2Z}. This offset applies to the 16 $\Gamma_\tau$-limited bursts, whose Lorentz factors are the largest in the sample, and it is one reason why our values for GRB~090926A and GRB~190114C exceed those obtained in time-resolved analyses of the same bursts.
It does not affect the 61 $\Gamma_{\max}$-limited bursts, for which the constraint is purely kinematic.
The adopted MVT introduces an additional dependence within the same calculation because $\delta t$ enters both $\Gamma_\tau$ and $R$.
For the $\Gamma_\tau$-limited GRBs, a shorter MVT gives a larger $\Gamma_\tau$, whereas $\Gamma_{\max}$ is independent of $\delta t$.
The emission radius depends on the MVT for all 77 GRBs. For the $\Gamma_{\max}$-limited sources, $R$ increases directly with $\delta t$, whereas for the $\Gamma_\tau$-limited sources this increase is partly offset by the decrease of $\Gamma_\tau$ with increasing $\delta t$.
\begin{figure}[tbp]
\centering
\includegraphics[width=0.76\textwidth]
{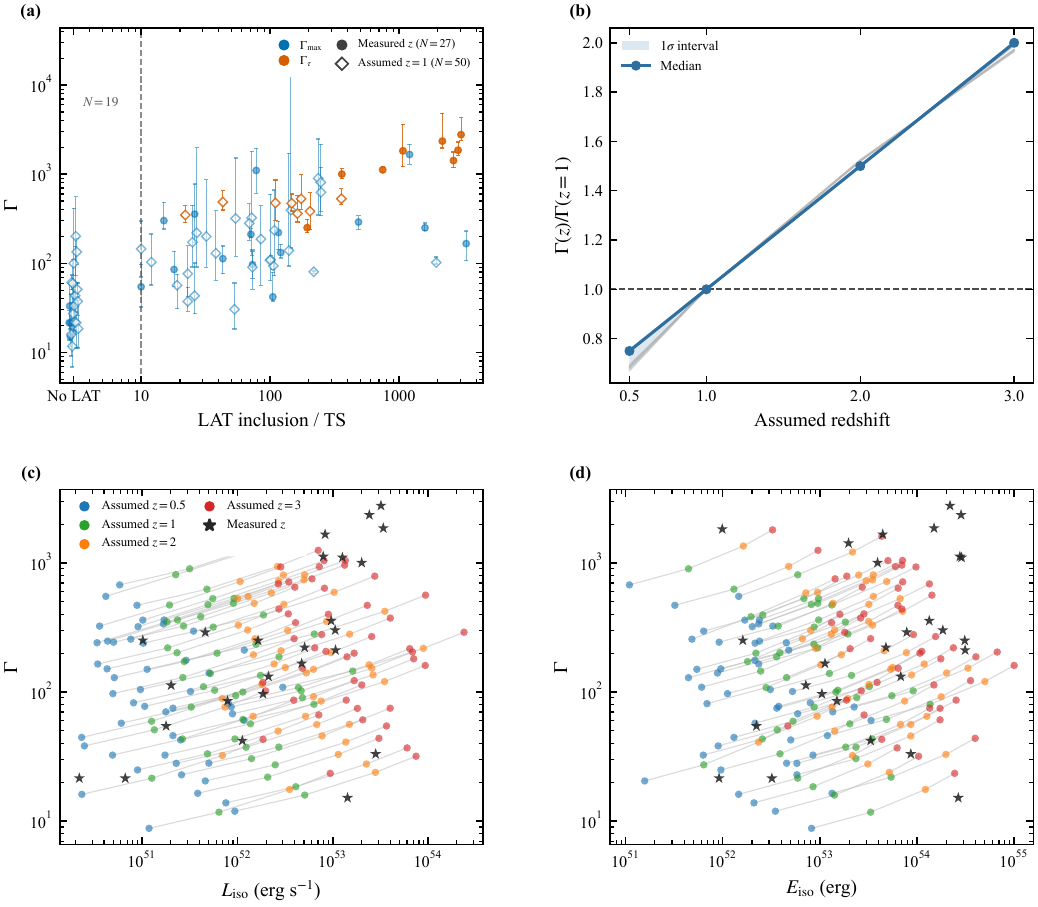}
\caption{LAT detectability and assumed-redshift effects on the inferred Lorentz factor and energetics.
Panel (a) compares the 19 GRBs fitted without LAT spectra with the 58 fitted with LAT spectra; for the latter, $\Gamma$ is plotted against LAT TS.
The dashed line marks the LAT inclusion threshold of $\mathrm{TS}=10$.
Colors denote the limiting constraint, and filled circles and open diamonds indicate measured and assumed redshifts, respectively.
Panel (b) shows $\Gamma(z)/\Gamma(z=1)$ for the 50 GRBs without measured redshifts; gray lines show individual GRBs, while the solid curve and shaded band give the median and $1\sigma$ interval.
Panels (c) and (d) show the corresponding trajectories in the $\Gamma$--$L_{\rm iso}$ and $\Gamma$--$E_{\rm iso}$ planes, with black stars denoting the measured-redshift sample.}
\label{fig:systematics}
\end{figure}

The GRBs fitted with and without LAT spectra show different $E_c$ and $\Gamma$ distributions.
The 58 GRBs fitted with LAT spectra have median $\Gamma=243$ and $E_c=74.5~{\rm MeV}$, compared with $\Gamma=33$ and $E_c=11.3~{\rm MeV}$ for the 19 GRBs fitted without LAT spectra.
This difference is consistent with LAT detectability, since GRBs with brighter high-energy emission and spectra extending to higher energies are more likely to be detected by the LAT.
The LAT-detected GRBs tend to have higher $E_c$ values and larger inferred $\Gamma$, as shown in panel~(a) of figure~\ref{fig:systematics}.
The assumed redshift also enters the Lorentz-factor calculation.
In particular, $\Gamma_{\max}$ scales as $(1+z)E_c$, while $\Gamma_\tau$ depends on redshift through both the explicit $(1+z)$ dependence and the luminosity distance.
Panel~(b) shows the effect of varying the assumed redshift from $z=0.5$ to 3.
Relative to $z=1$, the median $\Gamma$ changes by factors of approximately 0.75, 1.50, and 2.00 at $z=0.5$, 2, and 3, respectively, and only one GRB changes its limiting constraint.
In contrast, $L_{\rm iso}$ and $E_{\rm iso}$ vary by more than an order of magnitude over the same redshift range, as shown in panels~(c) and (d).
Because the same assumed redshift also determines the energetics through the luminosity distance, varying $z$ changes both coordinates in the $\Gamma$--energetics planes.
The full-sample $\Gamma$--energetics relations should therefore be interpreted with caution, whereas the $E_{c,z}$--$E_{p,z}$ analysis uses only GRBs with measured redshifts and is unaffected by the adopted $z=1$ for the remaining GRBs.

\section{Conclusions} \label{sec:conclusions}

We have carried out a systematic study of high-energy spectral cutoffs in the time-integrated prompt spectra of 139 \textit{Fermi} GRBs observed between 2008 July and 2025 December.
Our main results are as follows.

\begin{enumerate}
\item Joint GBM, LLE, and LAT fits with four empirical models yield 106 bursts with well-constrained cutoff energies.
A two-component Gaussian mixture model applied to $\log_{10}(E_c/{\rm MeV})$, with the measurement uncertainties propagated through Monte Carlo simulations, places the boundary between the two components at $E_c=2.12_{-0.21}^{+0.20}$~MeV.
The 77 bursts with $P_{\rm high}>0.99$ define the sample used for the pair-opacity analysis.

\item Combining the cutoff energies with the wavelet-based MVTs, we obtain bulk Lorentz factors of $11.7$--$2.78\times10^3$ (median 145) and emission radii spanning $\sim10^{13}$--$10^{15}~{\rm cm}$.
For 61 of the 77 bursts the constraint is set by the kinematic limit $\Gamma_{\max}$ rather than by $\tau_{\gamma\gamma}=1$.

\item No statistically significant $\Gamma$--$L_{\rm iso}$ or $\Gamma$--$E_{\rm iso}$ correlation is found in the full sample.
For the 27 GRBs with measured redshifts, the $\Gamma$--$L_{\rm iso}$ relation is moderate and statistically significant, with a slope consistent with previous $\Gamma_0$--$L_{\gamma,\rm iso}$ relation obtained independently from afterglow onset times, whereas the $\Gamma$--$E_{\rm iso}$ relation remains weak.

\item The strongest correlation found in this work is the rest-frame $E_{c,z}$--$E_{p,z}$ relation for the 26 GRBs with measured redshifts, with a Spearman coefficient of $\rho=0.72$ and a partial coefficient of $\rho=0.63$ after controlling for redshift.
The fitted slope of $2.15_{-0.97}^{+0.96}$ suggests a relation steeper than simple proportionality.
Within the internal pair-opacity interpretation, this relation may arise through the prompt photon field together with the bulk Lorentz factor and emission radius, although the detailed physical connection and theoretical scaling are not yet established.
\end{enumerate}

These results rest on time-integrated spectra, on a one-zone treatment of the pair opacity, and, for most of the sample, on an assumed redshift.
Larger samples with measured redshifts, time-resolved broadband spectroscopy,
and multi-zone opacity calculations will be needed to test the
$E_{c,z}$--$E_{p,z}$ relation and to determine how accurately the inferred
Lorentz factors and emission radii represent the properties of individual outflows.

\acknowledgments

This work made use of public data from the \textit{Fermi} Science Support Center and of the
\textit{Fermi}-GBM Burst Catalog maintained by the High Energy Astrophysics Science Archive Research
Center (HEASARC) at NASA/GSFC, and of the GRBweb compilation hosted by the IceCube Collaboration.

This work is based on observations with the \textit{Fermi} Gamma-ray Space Telescope (GBM and LAT).
The analysis made use of \texttt{GTBURST} (\textit{Fermi} Science Tools),
\texttt{XSPEC}~\cite{1996ASPC..101...17A},
\texttt{Astropy}~\cite{2018AJ....156..123A},
\texttt{NumPy}, \texttt{SciPy} and \texttt{Matplotlib}.

\input{Results/2026-08-01_v03_T90-1sigma-corrected/tables/high_ec_result_tables_t90_corrected.tex}
\appendix
\section{Empirical classification of the cutoff energies}
\label{app:ec-selection}

For the 106 GRBs with valid cutoff-energy measurements, we define $x_i=\log_{10}(E_{c,i}/{\rm MeV})$ and fit the distribution with a two-component Gaussian mixture model,
\begin{equation}
p(x)
=
\sum_{k=1}^{2}
\frac{w_k}{\sqrt{2\pi}\sigma_k}
\exp\left[-\frac{(x-\mu_k)^2}{2\sigma_k^2}\right],
\qquad
\sum_{k=1}^{2}w_k=1,
\label{eq:ec-k2-gmm}
\end{equation}
where $w_k$, $\mu_k$, and $\sigma_k$ are the weight,
mean,
and standard deviation of component $k$ in $x$,
respectively.
The components are ordered by their means,
$\mu_1<\mu_2$.
Using the best-fit $E_c$ values,
the lower- and higher-mean components have
$(w,\mu,\sigma)=(0.104,0.118,0.175)$ and
$(0.896,1.526,0.943)$,
respectively.
The weighted-density intersection between their means is
$E_c=1.889~{\rm MeV}$.
The fit to the best-fit $E_c$ values is shown in figure~\ref{fig:ec-k2}.
The adopted boundary is determined from the Monte Carlo analysis below.

\begin{figure}[!htbp]
\centering
\includegraphics[width=0.80\textwidth]
{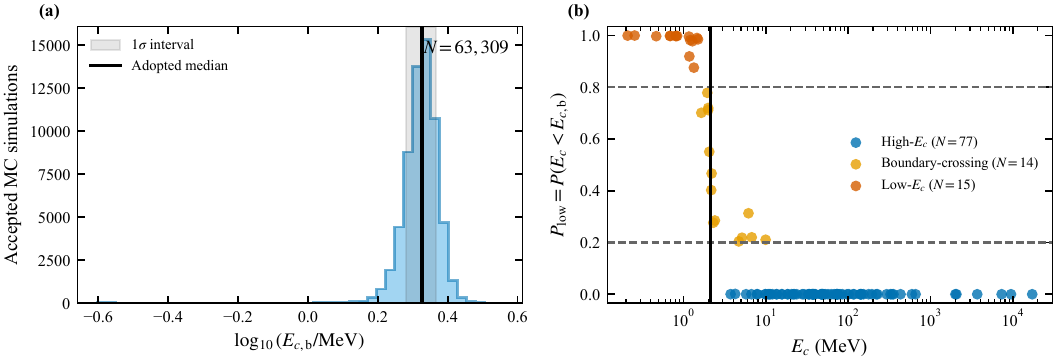}
\caption{Monte Carlo classification of the cutoff-energy distribution.
Panel (a) shows the distribution of the 63,309 retained intersections between the lower- and higher-mean Gaussian components.
The black line marks the adopted median boundary, and the gray interval denotes its $1\sigma$ uncertainty.
Panel (b) shows the low-$E_c$ probability $P_{\rm low}$ versus the best-fit $E_c$ for the 106 GRBs.
The dashed horizontal lines mark the probability thresholds separating the high-$E_c$, boundary-crossing, and low-$E_c$ groups, and the vertical black line marks the adopted cutoff-energy boundary at $E_c=2.120~{\rm MeV}$.}
\label{fig:ec-k2-mc}
\end{figure}

To propagate the asymmetric $E_c$ uncertainties,
we performed 100,000 Monte Carlo simulations.
In each simulation,
every $E_c$ was resampled in logarithmic space of its confidence interval,
and the two-component model was refitted.
After ordering the component means,
the candidate intersection $x_{\rm b}$ in logarithmic space was defined by
\begin{equation}
w_1\mathcal{N}(x_{\rm b}\mid\mu_1,\sigma_1^2)
=
w_2\mathcal{N}(x_{\rm b}\mid\mu_2,\sigma_2^2),
\label{eq:ec-k2-boundary}
\end{equation}
where $\mathcal{N}(x_{\rm b}\mid\mu,\sigma^2)$ denotes a Gaussian density
evaluated at $x_{\rm b}$,
with mean $\mu$ and variance $\sigma^2$.
The cutoff-energy boundary is
$E_{c,{\rm b}}=10^{x_{\rm b}}~{\rm MeV}$.
A simulation was retained only when this intersection was finite and lay
between the two component means,
and when the lower-mean component contained at least three GRBs but no more
than half of the sample.
This criterion avoids solutions driven by only one or two bursts,
as well as those in which the higher-mean component isolates only the extreme
high-$E_c$ tail.
These criteria retain 63,309 of the 100,000 simulations.
We adopt the median of the retained $x_{\rm b}$ distribution,
$x_{\rm b}=0.326_{-0.045}^{+0.040}$,
corresponding to
$E_{c,{\rm b}}=2.120_{-0.211}^{+0.203}~{\rm MeV}$.
The uncertainties represent the $1\sigma$ interval.
The retained boundary distribution is shown in
figure~\ref{fig:ec-k2-mc}(a).

For each GRB,
we then performed 100,000 paired draws from the $E_{c,i}$ and
$E_{c,{\rm b}}$ distributions to estimate the probabilities of lying below
and above the classification boundary:
\begin{equation}
P_{{\rm low},i}
=
\frac{N_{{\rm low},i}}{N},
\qquad
P_{{\rm high},i}=1-P_{{\rm low},i}.
\label{eq:ec-k2-membership}
\end{equation}
Here $N=100,000$,
and $N_{{\rm low},i}$ is the number of joint draws satisfying
$E_{c,i}^{(j)}<E_{c,{\rm b}}^{(j)}$.
The quantities $E_{c,i}^{(j)}$ and $E_{c,{\rm b}}^{(j)}$ are drawn from the
burst-specific cutoff-energy distribution and the retained boundary
distribution,
respectively.
GRBs with $P_{\rm low}<0.2$ are classified as high-$E_c$,
those with $0.2\leq P_{\rm low}\leq0.8$ as boundary-crossing,
and those with $P_{\rm low}>0.8$ as low-$E_c$.
This gives 77 high-$E_c$ bursts,
14 boundary-crossing bursts,
and 15 low-$E_c$ bursts.
Although the high-$E_c$ criterion requires only $P_{\rm high}>0.8$,
all 77 selected GRBs have $P_{\rm high}>0.99$,
with a minimum value of 0.99869.
The GRB classification is shown in
figure~\ref{fig:ec-k2-mc}(b).
Table~\ref{tab:ec-nonhigh-properties} lists the redshift,
$T_{90}$ interval,
best-fit spectral model and parameters,
fit statistic,
and $P_{\rm low}$ for the 29 GRBs not included in the high-$E_c$ selection.
The spectral parameter uncertainties correspond to 90\% confidence
intervals.

\begin{landscape}
\begingroup
\scriptsize
\setlength{\tabcolsep}{1pt}
\setlength{\LTcapwidth}{\linewidth}
\begin{longtable}{@{}lcclccccccccc@{}}
\caption{Burst and spectral-fit results of the
boundary-crossing and low-$E_c$ GRBs}\label{tab:ec-nonhigh-properties}\\
\toprule
GRB & $z$ & $T_{90}$ interval & Model & $\alpha$ & $\beta$ & $E_0$ & $N_0$ &
$\lambda$ & $E_c$ & $N_1$ & Stat./dof & $P_{\rm low}$ \\
 &  & (s) &  &  &  & (keV) &  &  & (MeV) &  &  &  \\
\midrule
\endfirsthead
\multicolumn{13}{c}{\tablename~\thetable{} (continued)}\\
\toprule
GRB & $z$ & $T_{90}$ interval & Model & $\alpha$ & $\beta$ & $E_0$ & $N_0$ &
$\lambda$ & $E_c$ & $N_1$ & Stat./dof & $P_{\rm low}$ \\
 &  & (s) &  &  &  & (keV) &  &  & (MeV) &  &  &  \\
\midrule
\endhead
\endfoot
\multicolumn{13}{c}{Boundary-crossing GRBs ($0.2\leq P_{\rm low}\leq0.8$)} \\
\midrule
081024B & $\cdots$ & $-0.06$--$0.58$ & CPL$_{\rm piv}$ & $\cdots$ & $\cdots$ & $\cdots$ & $\cdots$ & $-1.08_{-0.15}^{+0.30}$ & $4.68_{-3.66}^{+2.93}$ & $5.43_{-1.55}^{+6.26}$ & $360.9/407$ & $0.20419$ \\
110824A & $\cdots$ & $0.0$--$76.6$ & CPL$_{\rm piv}$ & $\cdots$ & $\cdots$ & $\cdots$ & $\cdots$ & $-1.34_{-0.07}^{+0.12}$ & $1.64_{-0.89}^{+1.62}$ & $1.37_{-0.27}^{+0.61}$ & $242.1/337$ & $0.70126$ \\
140306A & $\cdots$ & $1.1$--$52.8$ & CPL$_{\rm piv}$ & $\cdots$ & $\cdots$ & $\cdots$ & $\cdots$ & $-1.20_{-0.02}^{+0.02}$ & $1.99_{-0.33}^{+0.25}$ & $8.39_{-0.43}^{+0.49}$ & $315.7/336$ & $0.71171$ \\
140329A & $\cdots$ & $5.6$--$26.9$ & Band+cutoff & $-0.66_{-0.04}^{+0.05}$ & $-1.84_{-0.14}^{+0.20}$ & $144_{-16}^{+14}$ & $2.11_{-0.32}^{+0.31}$ & $\cdots$ & $2.19_{-1.14}^{+2.26}$ & $\cdots$ & $349.0/333$ & $0.46677$ \\
150403A & $2.06$ & $3.3$--$25.6$ & Band+cutoff & $-0.70_{-0.05}^{+0.06}$ & $-1.54_{-0.10}^{+0.09}$ & $211_{-31}^{+31}$ & $1.32_{-0.25}^{+0.21}$ & $\cdots$ & $2.40_{-0.68}^{+1.07}$ & $\cdots$ & $371.4/364$ & $0.28504$ \\
150820A & $\cdots$ & $-0.77$--$5.12$ & CPL$_{\rm piv}$ & $\cdots$ & $\cdots$ & $\cdots$ & $\cdots$ & $-1.52_{-0.24}^{+2.12}$ & $6.14_{-5.91}^{+15.12}$ & $0.40_{-0.21}^{+67.81}$ & $342.9/379$ & $0.31298$ \\
160101B & $\cdots$ & $0.0$--$22.0$ & CPL$_{\rm piv}$ & $\cdots$ & $\cdots$ & $\cdots$ & $\cdots$ & $-1.56_{-0.15}^{+0.25}$ & $9.92_{-9.44}^{+13.01}$ & $0.46_{-0.18}^{+0.66}$ & $340.1/379$ & $0.21069$ \\
170409A & $\cdots$ & $29.4$--$93.4$ & Band+cutoff & $-0.77_{-0.02}^{+0.00}$ & $-1.30_{-0.19}^{+0.12}$ & $623_{-118}^{+60}$ & $1.46_{-0.01}^{+0.00}$ & $\cdots$ & $1.94_{-0.13}^{+0.16}$ & $\cdots$ & $527.5/336$ & $0.77860$ \\
171227A & $\cdots$ & $7.7$--$45.4$ & Band+cutoff & $-0.61_{-0.01}^{+0.00}$ & $-1.22_{-0.03}^{+0.04}$ & $265_{-45}^{+1}$ & $1.70_{-0.17}^{+0.10}$ & $\cdots$ & $1.99_{-0.12}^{+0.12}$ & $\cdots$ & $554.7/334$ & $0.72004$ \\
180113C & $\cdots$ & $5.4$--$30.0$ & Band+cutoff & $-0.47_{-0.03}^{+0.04}$ & $-1.59_{-0.06}^{+0.05}$ & $143_{-11}^{+10}$ & $1.50_{-0.16}^{+0.18}$ & $\cdots$ & $2.30_{-0.38}^{+0.63}$ & $\cdots$ & $550.3/367$ & $0.27632$ \\
190415A & $\cdots$ & $0.8$--$53.3$ & CPL+CPL$_{\rm piv}$ & $1.84_{-1.19}^{+0.69}$ & $\cdots$ & $204_{-80}^{+376}$ & $0.00_{-0.00}^{+0.00}$ & $-1.24_{-0.11}^{+0.07}$ & $2.05_{-0.59}^{+0.72}$ & $5.46_{-2.24}^{+1.61}$ & $265.1/333$ & $0.55067$ \\
211211A & $0.108$ & $2.6$--$36.9$ & Band+cutoff & $-0.68_{-0.06}^{+0.07}$ & $-1.44_{-0.01}^{+0.01}$ & $87_{-12}^{+15}$ & $11.60_{-1.74}^{+1.79}$ & $\cdots$ & $2.17_{-0.10}^{+0.13}$ & $\cdots$ & $618.1/335$ & $0.40246$ \\
220210A & $\cdots$ & $-0.5$--$43.0$ & CPL$_{\rm piv}$ & $\cdots$ & $\cdots$ & $\cdots$ & $\cdots$ & $-1.37_{-0.20}^{+0.28}$ & $6.73_{-6.07}^{+6.14}$ & $0.46_{-0.20}^{+0.62}$ & $322.8/409$ & $0.21998$ \\
250620C & $\cdots$ & $0.0$--$31.8$ & CPL+CPL$_{\rm piv}$ & $-0.90_{-0.58}^{+0.89}$ & $\cdots$ & $301_{-198}^{+28}$ & $2.22_{-2.05}^{+0.92}$ & $-1.26_{-0.23}^{+1.21}$ & $5.12_{-4.25}^{+22.32}$ & $1.57_{-1.28}^{+12.50}$ & $347.8/334$ & $0.21849$ \\
\midrule
\multicolumn{13}{c}{Low-$E_c$ GRBs ($P_{\rm low}>0.8$)} \\
\midrule
100719D & $\cdots$ & $1.5$--$23.4$ & Band+cutoff & $-0.57_{-0.07}^{+0.07}$ & $-1.61_{-0.26}^{+0.18}$ & $165_{-20}^{+28}$ & $0.83_{-0.17}^{+0.22}$ & $\cdots$ & $1.18_{-0.43}^{+1.23}$ & $\cdots$ & $403.4/333$ & $0.91931$ \\
121225B & $\cdots$ & $9.5$--$68.0$ & CPL+CPL$_{\rm piv}$ & $0.17_{-0.76}^{+1.33}$ & $\cdots$ & $66_{-35}^{+56}$ & $0.01_{-0.01}^{+0.29}$ & $-1.35_{-0.09}^{+0.06}$ & $1.34_{-0.53}^{+1.20}$ & $5.54_{-1.98}^{+2.19}$ & $310.6/365$ & $0.87583$ \\
130821A & $\cdots$ & $3.6$--$90.6$ & CPL+CPL$_{\rm piv}$ & $1.35_{-1.07}^{+0.63}$ & $\cdots$ & $38_{-12}^{+31}$ & $0.00_{-0.00}^{+0.00}$ & $-1.36_{-0.05}^{+0.08}$ & $1.20_{-0.44}^{+0.77}$ & $4.04_{-1.02}^{+1.46}$ & $358.5/395$ & $0.98227$ \\
140102A & $\cdots$ & $0.45$--$4.10$ & CPL+CPL$_{\rm piv}$ & $0.24_{-0.72}^{+0.85}$ & $\cdots$ & $61_{-27}^{+51}$ & $0.07_{-0.07}^{+0.77}$ & $-1.21_{-0.22}^{+0.14}$ & $0.47_{-0.16}^{+1.23}$ & $38.56_{-28.92}^{+35.74}$ & $370.8/362$ & $0.99746$ \\
150510A & $\cdots$ & $0.4$--$52.3$ & CPL$_{\rm piv}$ & $\cdots$ & $\cdots$ & $\cdots$ & $\cdots$ & $-1.03_{-0.02}^{+0.02}$ & $1.46_{-0.17}^{+0.12}$ & $16.52_{-0.89}^{+1.07}$ & $375.3/408$ & $0.99052$ \\
150514A & $0.807$ & $0.0$--$10.8$ & CPL$_{\rm piv}$ & $\cdots$ & $\cdots$ & $\cdots$ & $\cdots$ & $-1.56_{-0.11}^{+0.12}$ & $0.21_{-0.06}^{+0.12}$ & $3.66_{-1.32}^{+2.44}$ & $409.6/378$ & $0.99969$ \\
170527A & $\cdots$ & $1.1$--$50.2$ & CPL$_{\rm piv}$ & $\cdots$ & $\cdots$ & $\cdots$ & $\cdots$ & $-1.09_{-0.02}^{+0.03}$ & $1.16_{-0.16}^{+0.10}$ & $14.21_{-0.75}^{+1.36}$ & $656.7/339$ & $0.99595$ \\
170808B & $\cdots$ & $4.1$--$21.8$ & Band+cutoff & $-0.87_{-0.02}^{+0.03}$ & $-1.72_{-0.14}^{+0.08}$ & $188_{-13}^{+14}$ & $9.44_{-0.78}^{+0.66}$ & $\cdots$ & $1.28_{-0.28}^{+0.70}$ & $\cdots$ & $412.0/365$ & $0.97797$ \\
171010A & $0.329$ & $16.6$--$123.9$ & CPL$_{\rm piv}$ & $\cdots$ & $\cdots$ & $\cdots$ & $\cdots$ & $-1.20_{-0.00}^{+0.00}$ & $0.25_{-0.01}^{+0.00}$ & $93.52_{-0.96}^{+0.32}$ & $992.0/335$ & $0.99999$ \\
181222A & $\cdots$ & $2.2$--$132.9$ & CPL$_{\rm piv}$ & $\cdots$ & $\cdots$ & $\cdots$ & $\cdots$ & $-1.08_{-0.08}^{+0.05}$ & $0.80_{-0.18}^{+0.42}$ & $14.42_{-3.07}^{+2.79}$ & $317.8/335$ & $0.99800$ \\
190727B & $\cdots$ & $3.1$--$38.1$ & CPL$_{\rm piv}$ & $\cdots$ & $\cdots$ & $\cdots$ & $\cdots$ & $-1.25_{-0.02}^{+0.03}$ & $0.84_{-0.15}^{+0.07}$ & $13.93_{-0.75}^{+1.76}$ & $342.7/377$ & $0.99839$ \\
200826B & $\cdots$ & $2.30$--$9.73$ & Band+cutoff & $-0.47_{-0.03}^{+0.03}$ & $-1.16_{-0.10}^{+0.07}$ & $205_{-16}^{+15}$ & $3.09_{-0.29}^{+0.29}$ & $\cdots$ & $0.68_{-0.07}^{+0.11}$ & $\cdots$ & $572.4/337$ & $0.99853$ \\
230812B & $0.36$ & $0.38$--$3.65$ & Band+cutoff & $-0.53_{-0.02}^{+0.02}$ & $-1.78_{-0.07}^{+0.08}$ & $129_{-5}^{+5}$ & $24.83_{-1.61}^{+1.60}$ & $\cdots$ & $0.81_{-0.10}^{+0.12}$ & $\cdots$ & $1597.3/408$ & $0.99839$ \\
231104A & $\cdots$ & $8.7$--$45.8$ & CPL$_{\rm piv}$ & $\cdots$ & $\cdots$ & $\cdots$ & $\cdots$ & $-1.03_{-0.03}^{+0.04}$ & $0.69_{-0.08}^{+0.09}$ & $15.44_{-1.51}^{+2.01}$ & $618.0/407$ & $0.99853$ \\
240718A & $\cdots$ & $0.3$--$58.9$ & CPL$_{\rm piv}$ & $\cdots$ & $\cdots$ & $\cdots$ & $\cdots$ & $-1.01_{-0.03}^{+0.05}$ & $1.50_{-0.20}^{+0.24}$ & $8.00_{-0.54}^{+0.59}$ & $311.1/335$ & $0.98527$ \\
\bottomrule
\end{longtable}
\endgroup
\end{landscape}

\section{\texorpdfstring{Additional prompt spectral fits}{Additional prompt spectral fits}}
\label{app:spectral-figures}
\begin{figure}[H]
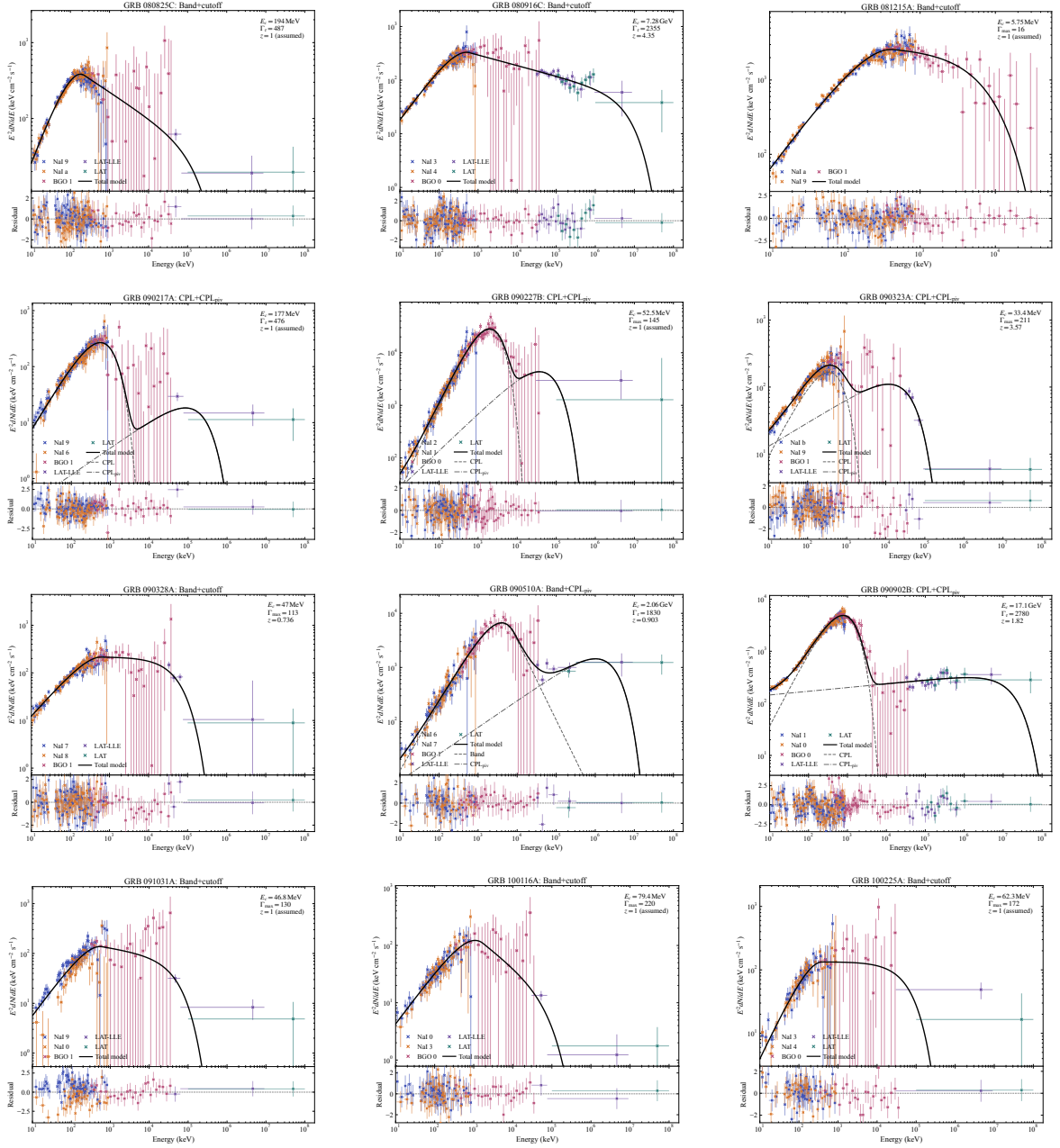

\renewcommand{\specfigwidth}{0.30\textwidth}
\centering
\specfig{080825C}\hfill\specfig{080916C}\hfill\specfig{081215A} \\[6pt]
\specfig{090217A}\hfill\specfig{090227B}\hfill\specfig{090323A} \\[6pt]
\specfig{090328A}\hfill\specfig{090510A}\hfill\specfig{090902B} \\[6pt]
\specfig{091031A}\hfill\specfig{100116A}\hfill\specfig{100225A}
\caption{Time-integrated prompt spectra and best-fit models for the 71 high-$E_c$ GRBs not shown in figure~\ref{fig:spectral-examples}. Symbols, curves and annotations follow figure~\ref{fig:spectral-examples}. GRBs~080825C--100225A.}
\label{fig:specset}
\end{figure}

Figure~\ref{fig:specset} shows the time-integrated $\nu F_\nu$ spectra and best-fit models for the remaining 71 GRBs in the pair-opacity sample, excluding the six examples in figure~\ref{fig:spectral-examples}.
The bursts are ordered by trigger date, as in table~\ref{tab:mvt-results}; spectral parameters and derived quantities are listed in table~\ref{spectra}.

\begin{figure}[p]
\centering
\specfig{100724B}\hfill\specfig{100826A}\hfill\specfig{101123A} \\[6pt]
\specfig{110328B}\hfill\specfig{110529A}\hfill\specfig{110721A} \\[6pt]
\specfig{110731A}\hfill\specfig{120328B}\hfill\specfig{120709A} \\[6pt]
\specfig{121011A}\hfill\specfig{130305A}\hfill\specfig{130310A}
\continuedcaption{Continued. GRBs~100724B--130310A.}
\end{figure}

\begin{figure}[p]
\centering
\specfig{130504C}\hfill\specfig{130518A}\hfill\specfig{130606B} \\[6pt]
\specfig{130828A}\hfill\specfig{131108A}\hfill\specfig{131216A} \\[6pt]
\specfig{131231A}\hfill\specfig{140110A}\hfill\specfig{140206B} \\[6pt]
\specfig{140619B}\hfill\specfig{141028A}\hfill\specfig{141207A}
\continuedcaption{Continued. GRBs~130504C--141207A.}
\end{figure}

\begin{figure}[p]
\centering
\specfig{141222A}\hfill\specfig{150210A}\hfill\specfig{150416A} \\[6pt]
\specfig{150523A}\hfill\specfig{151006A}\hfill\specfig{160509A} \\[6pt]
\specfig{160709A}\hfill\specfig{160816A}\hfill\specfig{160821A} \\[6pt]
\specfig{160905A}\hfill\specfig{160910A}\hfill\specfig{170214A}
\continuedcaption{Continued. GRBs~141222A--170214A.}
\end{figure}

\begin{figure}[p]
\centering
\specfig{170405A}\hfill\specfig{170424A}\hfill\specfig{180703A} \\[6pt]
\specfig{180720B}\hfill\specfig{190531B}\hfill\specfig{200101A} \\[6pt]
\specfig{200313A}\hfill\specfig{200412B}\hfill\specfig{200829A} \\[6pt]
\specfig{201104A}\hfill\specfig{210410A}\hfill\specfig{210619B}
\continuedcaption{Continued. GRBs~170405A--210619B.}
\end{figure}

\begin{figure}[p]
\centering
\specfig{220101A}\hfill\specfig{220107B}\hfill\specfig{220209A} \\[6pt]
\specfig{221023A}\hfill\specfig{221027A}\hfill\specfig{221209A} \\[6pt]
\specfig{230512A}\hfill\specfig{240118A}\hfill\specfig{240825A} \\[6pt]
\specfig{240905B}\hfill\specfig{250716A}\hfill\makebox[\specfigwidth]{}
\continuedcaption{Continued. GRBs~220101A--250716A.}
\end{figure}

\clearpage \input{jcapexample.bbl}

\end{document}

%% file: Results/2026-08-01_v03_T90-1sigma-corrected/tables/high_ec_result_tables_t90_corrected.tex
\clearpage
\begingroup
\scriptsize
\setlength{\tabcolsep}{3pt}
\setlength{\LTcapwidth}{\linewidth}
\begin{longtable}{@{}ccccccc@{}}
\caption{GRB information and MVT results for the 77 high-$E_c$ GRBs}\label{tab:mvt-results}\\
\toprule
GRB & 
  z & 
  T$_{90}$ interval (s) & 
  $\Delta T$ (s) & 
  $\delta t$ (s) &
  Detectors &
  LAT TS \\
\midrule
\endfirsthead
\multicolumn{7}{c}{\tablename~\thetable{} (continued)}\\
\toprule
GRB & 
  z & 
  T$_{90}$ interval (s) & 
  $\Delta T$ (s) & 
  $\delta t$ (s) &
  Detectors &
  LAT TS \\
\midrule
\endhead
\midrule
\multicolumn{7}{r}{Continued on next page}\\
\endfoot
\bottomrule
\endlastfoot
080825C & $\cdots$ & $1.2$--$22.2$ & $21.0_{-0.2}^{+0.2}$ & $0.0802_{-0.0098}^{+0.0111}$ & n9, na, b1, LLE, LAT & $43$ \\
080916C & $4.35$ & $1.3$--$64.3$ & $63.0_{-0.8}^{+0.8}$ & $0.2585_{-0.0561}^{+0.0716}$ & n3, n4, b0, LLE, LAT & $2169$ \\
081215A & $\cdots$ & $1.22$--$6.78$ & $5.57_{-0.14}^{+0.14}$ & $0.0248_{-0.0089}^{+0.0140}$ & na, n9, b1 & $\cdots$ \\
090217A & $\cdots$ & $0.8$--$34.1$ & $33.3_{-0.7}^{+0.7}$ & $0.1297_{-0.0204}^{+0.0242}$ & n9, n6, b1, LLE, LAT & $110$ \\
090227B & $\cdots$ & $-0.02$--$0.29$ & $0.30_{-0.02}^{+0.02}$ & $0.0025_{-0.0014}^{+0.0034}$ & n2, n1, b0, LLE, LAT & $10$ \\
090323A & $3.57$ & $8.7$--$142.6$ & $133.9_{-0.6}^{+0.6}$ & $0.5739_{-0.0261}^{+0.0274}$ & nb, n9, b1, LLE, LAT & $71$ \\
090328A & $0.736$ & $4.4$--$66.0$ & $61.7_{-1.8}^{+1.8}$ & $0.0759_{-0.0222}^{+0.0314}$ & n7, n8, b1, LLE, LAT & $43$ \\
090510A & $0.903$ & $-0.05$--$0.91$ & $0.96_{-0.14}^{+0.14}$ & $0.0059_{-0.0030}^{+0.0055}$ & n6, n7, b1, LLE, LAT & $1072$ \\
090902B & $1.82$ & $2.8$--$22.1$ & $19.3_{-0.3}^{+0.3}$ & $0.0316_{-0.0022}^{+0.0024}$ & n1, n0, b0, LLE, LAT & $3028$ \\
090926A & $2.11$ & $2.2$--$15.9$ & $13.8_{-0.3}^{+0.3}$ & $0.0801_{-0.0033}^{+0.0034}$ & n7, n6, b1, LLE, LAT & $2870$ \\
091031A & $\cdots$ & $1.4$--$35.3$ & $33.9_{-0.5}^{+0.5}$ & $0.0470_{-0.0160}^{+0.0243}$ & n9, n0, b1, LLE, LAT & $38$ \\
100116A & $\cdots$ & $0.6$--$103.1$ & $102.5_{-1.5}^{+1.5}$ & $0.2824_{-0.0368}^{+0.0424}$ & n0, n3, b0, LLE, LAT & $27$ \\
100225A & $\cdots$ & $-0.3$--$12.7$ & $13.0_{-1.9}^{+1.9}$ & $0.4322_{-0.1413}^{+0.2100}$ & n3, n4, b0, LLE, LAT & $25$ \\
100724B & $\cdots$ & $8.2$--$122.9$ & $114.7_{-3.2}^{+3.2}$ & $0.2321_{-0.0056}^{+0.0058}$ & n1, n0, b0, LLE, LAT & $106$ \\
100826A & $\cdots$ & $8.7$--$93.7$ & $85.0_{-0.7}^{+0.7}$ & $0.1821_{-0.0086}^{+0.0090}$ & n7, n8, b1, LLE & $\cdots$ \\
101123A & $\cdots$ & $41.5$--$145.4$ & $103.9_{-0.6}^{+0.6}$ & $0.1852_{-0.0140}^{+0.0152}$ & na, n9, b1 & $\cdots$ \\
110328B & $\cdots$ & $2.6$--$86.5$ & $84.0_{-3.5}^{+3.5}$ & $0.5015_{-0.3167}^{+0.8596}$ & n9, n0, b1, LLE & $\cdots$ \\
110529A & $\cdots$ & $-0.13$--$0.38$ & $0.51_{-0.09}^{+0.09}$ & $0.0010_{-0.0006}^{+0.0017}$ & n9, n6, b1, LLE & $\cdots$ \\
110721A & $0.382$ & $0.0$--$21.8$ & $21.8_{-0.6}^{+0.6}$ & $0.1917_{-0.0060}^{+0.0062}$ & n9, n6, b1, LLE, LAT & $194$ \\
110731A & $2.83$ & $0.00$--$7.49$ & $7.49_{-0.57}^{+0.57}$ & $0.1122_{-0.0059}^{+0.0062}$ & n0, n6, b0, LLE, LAT & $360$ \\
120226A & $\cdots$ & $4.4$--$57.3$ & $53.0_{-0.6}^{+0.6}$ & $0.2789_{-0.0406}^{+0.0476}$ & n1, n0, b0, LLE, LAT & $12$ \\
120328B & $\cdots$ & $3.8$--$33.5$ & $29.7_{-1.1}^{+1.1}$ & $0.2738_{-0.0061}^{+0.0062}$ & nb, n7, b1, LLE, LAT & $19$ \\
120709A & $\cdots$ & $-0.1$--$27.2$ & $27.3_{-1.0}^{+1.0}$ & $0.0430_{-0.0173}^{+0.0289}$ & n6, n7, b1, LLE, LAT & $72$ \\
121011A & $0.58$ & $1.0$--$66.8$ & $65.8_{-4.4}^{+4.4}$ & $0.0044_{-0.0028}^{+0.0074}$ & n1, n0, b0, LLE & $\cdots$ \\
130305A & $\cdots$ & $1.3$--$26.9$ & $25.6_{-1.6}^{+1.6}$ & $0.2694_{-0.0426}^{+0.0506}$ & n9, n6, b1, LLE & $\cdots$ \\
130310A & $\cdots$ & $4.1$--$20.1$ & $16.0_{-2.6}^{+2.6}$ & $0.0013_{-0.0004}^{+0.0006}$ & na, n9, b1, LLE & $\cdots$ \\
130504C & $\cdots$ & $8.7$--$81.9$ & $73.2_{-2.1}^{+2.1}$ & $0.1194_{-0.0066}^{+0.0070}$ & n9, n2, b0, LLE, LAT & $26$ \\
130518A & $2.49$ & $9.9$--$58.5$ & $48.6_{-0.9}^{+0.9}$ & $0.2677_{-0.0220}^{+0.0240}$ & n3, n6, b0, LLE, LAT & $26$ \\
130606B & $\cdots$ & $5.4$--$57.6$ & $52.2_{-0.7}^{+0.7}$ & $0.1622_{-0.0091}^{+0.0096}$ & n8, nb, b1 & $\cdots$ \\
130828A & $\cdots$ & $13.3$--$150.3$ & $137.0_{-0.9}^{+0.9}$ & $0.2052_{-0.0681}^{+0.1020}$ & n3, n0, b0, LLE, LAT & $147$ \\
131108A & $2.4$ & $0.3$--$18.5$ & $18.2_{-0.6}^{+0.6}$ & $0.0168_{-0.0030}^{+0.0037}$ & n6, n3, b0, LLE, LAT & $1207$ \\
131216A & $\cdots$ & $0.0$--$19.3$ & $19.3_{-3.6}^{+3.6}$ & $0.2158_{-0.0509}^{+0.0665}$ & n9, n6, b1, LLE & $\cdots$ \\
131231A & $0.642$ & $13.3$--$44.5$ & $31.2_{-0.6}^{+0.6}$ & $0.2624_{-0.0162}^{+0.0172}$ & n3, n6, b0, LLE, LAT & $18$ \\
140110A & $\cdots$ & $-0.26$--$9.22$ & $9.47_{-1.62}^{+1.62}$ & $0.2665_{-0.0410}^{+0.0485}$ & n7, n6, b1, LLE, LAT & $204$ \\
140206B & $\cdots$ & $7.5$--$154.2$ & $146.7_{-4.4}^{+4.4}$ & $0.3570_{-0.0217}^{+0.0231}$ & n1, n0, b0, LLE, LAT & $101$ \\
140619B & $\cdots$ & $-0.26$--$2.56$ & $2.82_{-0.81}^{+0.81}$ & $0.0170_{-0.0169}^{+2.4930}$ & n9, n0, b1, LLE, LAT & $236$ \\
141028A & $2.33$ & $6.7$--$38.1$ & $31.5_{-2.4}^{+2.4}$ & $0.3410_{-0.0431}^{+0.0494}$ & n9, n6, b1, LLE, LAT & $116$ \\
141207A & $\cdots$ & $1.3$--$22.3$ & $21.0_{-0.6}^{+0.6}$ & $0.2639_{-0.0386}^{+0.0453}$ & n3, n4, b0, LLE, LAT & $357$ \\
141222A & $\cdots$ & $0.00$--$2.75$ & $2.75_{-0.26}^{+0.26}$ & $0.0026_{-0.0005}^{+0.0005}$ & n3, n0, b0, LLE, LAT & $145$ \\
150210A & $\cdots$ & $0.0$--$31.3$ & $31.3_{-0.6}^{+0.6}$ & $0.0272_{-0.0062}^{+0.0080}$ & n9, na, b1, LLE, LAT & $54$ \\
150416A & $\cdots$ & $0.5$--$33.8$ & $33.3_{-3.7}^{+3.7}$ & $0.7904_{-0.1540}^{+0.1913}$ & n2, n1, b0, LLE & $\cdots$ \\
150523A & $\cdots$ & $1.8$--$84.2$ & $82.4_{-2.6}^{+2.6}$ & $0.3734_{-0.0835}^{+0.1075}$ & n0, n1, b0, LLE, LAT & $163$ \\
151006A & $\cdots$ & $1.5$--$95.0$ & $93.4_{-6.1}^{+6.1}$ & $0.9802_{-0.2184}^{+0.2811}$ & n0, n3, b0, LLE & $\cdots$ \\
160509A & $1.17$ & $8.2$--$377.9$ & $369.7_{-0.8}^{+0.8}$ & $0.1561_{-0.0256}^{+0.0306}$ & n3, n0, b0, LLE, LAT & $485$ \\
160625B & $1.41$ & $188.5$--$641.8$ & $453.4_{-0.6}^{+0.6}$ & $0.1092_{-0.0065}^{+0.0069}$ & n9, n6, b1, LLE, LAT & $1590$ \\
160709A & $\cdots$ & $0.00$--$5.44$ & $5.44_{-0.45}^{+0.45}$ & $0.0052_{-0.0042}^{+0.0202}$ & n3, n4, b0, LLE, LAT & $248$ \\
160816A & $\cdots$ & $0.4$--$11.5$ & $11.1_{-0.1}^{+0.1}$ & $0.0143_{-0.0028}^{+0.0034}$ & n6, n7, b1, LLE, LAT & $108$ \\
160821A & $\cdots$ & $118.5$--$161.5$ & $43.0_{-0.7}^{+0.7}$ & $0.2174_{-0.0359}^{+0.0430}$ & n6, n7, b1, LLE, LAT & $218$ \\
160905A & $\cdots$ & $3.8$--$37.4$ & $33.5_{-1.4}^{+1.4}$ & $0.2752_{-0.0280}^{+0.0311}$ & n0, n6, b1, LLE, LAT & $53$ \\
160910A & $\cdots$ & $4.6$--$28.9$ & $24.3_{-0.4}^{+0.4}$ & $0.1334_{-0.0087}^{+0.0093}$ & n5, n1, b0, LLE & $\cdots$ \\
170214A & $2.53$ & $12.5$--$135.4$ & $122.9_{-0.7}^{+0.7}$ & $0.1119_{-0.0106}^{+0.0117}$ & n0, n1, b0, LLE, LAT & $747$ \\
170405A & $3.51$ & $7.4$--$86.0$ & $78.6_{-0.6}^{+0.6}$ & $0.2530_{-0.0301}^{+0.0342}$ & n7, n6, b1, LLE, LAT & $15$ \\
170424A & $\cdots$ & $2.8$--$56.1$ & $53.2_{-3.2}^{+3.2}$ & $0.0756_{-0.0637}^{+0.4038}$ & n1, n0, b0, LLE, LAT & $23$ \\
180703A & $0.668$ & $1.5$--$22.3$ & $20.7_{-1.6}^{+1.6}$ & $0.4034_{-0.0433}^{+0.0485}$ & n7, n6, b1, LLE, LAT & $10$ \\
180720B & $0.654$ & $4.4$--$53.2$ & $48.9_{-0.4}^{+0.4}$ & $0.0246_{-0.0067}^{+0.0092}$ & n7, n6, b1, LLE, LAT & $105$ \\
190114C & $0.42$ & $0.7$--$117.1$ & $116.4_{-2.6}^{+2.6}$ & $0.0293_{-0.0014}^{+0.0015}$ & n4, n3, b0, LLE, LAT & $2644$ \\
190122A & $\cdots$ & $0.0$--$54.0$ & $54.0_{-3.3}^{+3.3}$ & $0.2160_{-0.0329}^{+0.0388}$ & n6, n9, b1, LLE & $\cdots$ \\
190531B & $\cdots$ & $6.7$--$44.9$ & $38.1_{-0.9}^{+0.9}$ & $0.1117_{-0.0064}^{+0.0068}$ & n3, n6, b0, LLE, LAT & $23$ \\
200101A & $\cdots$ & $4.9$--$14.9$ & $9.98_{-0.33}^{+0.33}$ & $0.0698_{-0.0077}^{+0.0086}$ & n5, n1, b0, LAT & $73$ \\
200313A & $\cdots$ & $0.8$--$14.3$ & $13.6_{-0.6}^{+0.6}$ & $0.1625_{-0.0053}^{+0.0055}$ & n8, n4, b1 & $\cdots$ \\
200412B & $\cdots$ & $5.2$--$11.3$ & $6.08_{-0.29}^{+0.29}$ & $0.0508_{-0.0030}^{+0.0032}$ & n7, n8, b1, LLE, LAT & $69$ \\
200829A & $1.25$ & $16.9$--$23.8$ & $6.91_{-0.36}^{+0.36}$ & $0.1467_{-0.0127}^{+0.0140}$ & n4, n8, b0 & $\cdots$ \\
201104A & $\cdots$ & $0.0$--$52.5$ & $52.5_{-7.4}^{+7.4}$ & $0.0665_{-0.0154}^{+0.0200}$ & n7, n3, b1, LLE, LAT & $247$ \\
210410A & $\cdots$ & $0.0$--$48.1$ & $48.1_{-2.8}^{+2.8}$ & $0.0890_{-0.0298}^{+0.0449}$ & n7, n6, b1, LAT & $140$ \\
210619B & $1.94$ & $0.6$--$55.4$ & $54.8_{-0.6}^{+0.6}$ & $0.0867_{-0.0043}^{+0.0046}$ & n8, n4, b1 & $\cdots$ \\
220101A & $4.62$ & $17.5$--$145.7$ & $128.3_{-15.8}^{+15.8}$ & $0.0596_{-0.0416}^{+0.1383}$ & n0, n7, b1, LLE, LAT & $78$ \\
220107B & $\cdots$ & $0.8$--$25.7$ & $24.8_{-0.4}^{+0.4}$ & $0.1038_{-0.0171}^{+0.0205}$ & n9, n1, b1, LLE, LAT & $174$ \\
220209A & $\cdots$ & $10.0$--$210.2$ & $200.2_{-0.4}^{+0.4}$ & $0.4051_{-0.0864}^{+0.1098}$ & n1, n2, b0, LLE, LAT & $22$ \\
220527A & $0.857$ & $2.6$--$13.1$ & $10.5_{-0.4}^{+0.4}$ & $0.0573_{-0.0043}^{+0.0046}$ & n7, n6, b1, LLE, LAT & $73$ \\
221023A & $\cdots$ & $3.1$--$42.3$ & $39.2_{-0.7}^{+0.7}$ & $0.1604_{-0.0173}^{+0.0193}$ & n1, n0, b0, LAT & $1944$ \\
221027A & $\cdots$ & $0.26$--$7.42$ & $7.17_{-0.72}^{+0.72}$ & $0.1186_{-0.0371}^{+0.0541}$ & n3, n4, b0, LLE, LAT & $32$ \\
221209A & $\cdots$ & $1.02$--$5.18$ & $4.16_{-0.83}^{+0.83}$ & $0.0797_{-0.0131}^{+0.0157}$ & n2, n5, b0 & $\cdots$ \\
230512A & $\cdots$ & $0.00$--$2.62$ & $2.62_{-0.09}^{+0.09}$ & $0.0147_{-0.0076}^{+0.0157}$ & n7, n6, b1, LLE, LAT & $85$ \\
240118A & $0.68$ & $45.3$--$99.3$ & $54.0_{-0.4}^{+0.4}$ & $0.0620_{-0.0025}^{+0.0026}$ & na, n2, b1, LLE, LAT & $121$ \\
240825A & $0.659$ & $1.15$--$5.12$ & $3.97_{-0.09}^{+0.09}$ & $0.0163_{-0.0019}^{+0.0022}$ & n7, n6, b1, LAT & $3316$ \\
240905B & $\cdots$ & $0.00$--$1.28$ & $1.28_{-0.20}^{+0.20}$ & $0.0162_{-0.0026}^{+0.0032}$ & n9, n6, b1, LAT & $100$ \\
250716A & $0.374$ & $4.4$--$70.4$ & $66.0_{-0.7}^{+0.7}$ & $0.3950_{-0.0181}^{+0.0190}$ & n4, n5, b0 & $\cdots$ \\
\end{longtable}
\noindent\footnotesize The detector column lists the data used in the joint spectral fit.
An ellipsis in the LAT TS column indicates that LAT data were not included.
For GRBs without a measured redshift,
$z=1$ is adopted in the pair-opacity calculation.
The uncertainties of $\delta t$ and $\Delta T$ are 90\% and catalog
1$\sigma$ intervals, respectively.
\par\endgroup

\begin{landscape}
\begingroup
\scriptsize
\setlength{\tabcolsep}{1pt}
\setlength{\LTcapwidth}{\linewidth}
\begin{longtable}{@{}ccccccccccccc@{}}
\caption{Spectral-fit and pair-opacity results for the 77 high-$E_c$ GRBs}\label{spectra}\\
\toprule
GRB &
  Model &
  $\alpha$ &
  $\beta$ &
  $E_0$ &
  $N_0$ &
  $\lambda$ &
  $E_c$ &
  $N_1$ &
  Stat./dof &
  $\Gamma$ &
  $E_{\rm ann}$ &
  $R$
  \\
   &
   &
   &
   &
  (keV) &
   &
   &
  (MeV) &
   &
   &
   &
  (MeV) &
  ($10^{14}~{\rm cm}$) \\
\midrule
\endfirsthead
\multicolumn{13}{c}{\tablename~\thetable{} (continued)}\\
\toprule
GRB &
  Model &
  $\alpha$ &
  $\beta$ &
  $E_0$ &
  $N_0$ &
  $\lambda$ &
  $E_c$ &
  $N_1$ &
  Stat./dof &
  $\Gamma$ &
  $E_{\rm ann}$ &
  $R$
  \\
   &
   &
   &
   &
  (keV) &
   &
   &
  (MeV) &
   &
   &
   &
  (MeV) &
  ($10^{14}~{\rm cm}$) \\
\midrule
\endhead
\midrule
\multicolumn{13}{r}{Continued on next page}\\
\endfoot
\bottomrule
\endlastfoot
080825C & Band+cutoff & $-0.62_{-0.06}^{+0.08}$ & $-2.34_{-0.09}^{+0.09}$ & $132_{-15}^{+15}$ & $1.16_{-0.26}^{+0.27}$ & $\cdots$ & $193.63_{-107.63}^{+528.03}$ & $\cdots$ & $311.5/403$ & $487_{-89}^{+167}$ & $159.85_{-90.07}^{+115.73}$ & $2.85_{-0.94}^{+2.3}$ \\
080916C & Band+cutoff & $-1.01_{-0.04}^{+0.04}$ & $-2.20_{-0.02}^{+0.02}$ & $510_{-71}^{+83}$ & $1.91_{-0.23}^{+0.26}$ & $\cdots$ & $7284.53_{-4117.64}^{+306610.16}$ & $\cdots$ & $352.5/405$ & $2355_{-387}^{+2464}$ & $13.89_{-12.60}^{+10.00}$ & $80.3_{-23}^{+254}$ \\
081215A & Band+cutoff & $-0.64_{-0.03}^{+0.04}$ & $-2.02_{-0.19}^{+0.19}$ & $301_{-30}^{+29}$ & $2.84_{-0.34}^{+0.37}$ & $\cdots$ & $5.75_{-2.46}^{+7.42}$ & $\cdots$ & $348.3/332$ & $15.9_{-6.8}^{+20.4}$ & $5.75_{-2.45}^{+7.36}$ & $(9.41_{-6.6}^{+44})\times10^{-4}$ \\
090217A & CPL+CPL$_{\rm piv}$ & $-0.83_{-0.06}^{+0.09}$ & $\cdots$ & $472_{-74}^{+80}$ & $0.52_{-0.17}^{+0.16}$ & $-1.52_{-0.23}^{+2.69}$ & $176.69_{-151.13}^{+391.33}$ & $0.04_{-0.03}^{+0.08}$ & $380.7/395$ & $476_{-173}^{+389}$ & $167.44_{-120.90}^{+617.38}$ & $4.40_{-2.6}^{+10}$ \\
090227B & CPL+CPL$_{\rm piv}$ & $-0.34_{-0.09}^{+0.10}$ & $\cdots$ & $1171_{-142}^{+156}$ & $0.52_{-0.23}^{+0.37}$ & $-1.30_{-0.11}^{+0.17}$ & $52.50_{-27.18}^{+55.14}$ & $6.93_{-3.36}^{+5.30}$ & $384.2/396$ & $145_{-75}^{+153}$ & $52.50_{-27.11}^{+55.23}$ & $(7.94_{-6.5}^{+37})\times10^{-3}$ \\
090323A & CPL+CPL$_{\rm piv}$ & $-0.88_{-0.21}^{+0.02}$ & $\cdots$ & $309_{-53}^{+113}$ & $0.72_{-0.04}^{+0.85}$ & $-1.65_{-0.03}^{+0.20}$ & $33.39_{-2.40}^{+5.42}$ & $0.66_{-0.14}^{+0.03}$ & $340.5/394$ & $211_{-15}^{+34}$ & $33.39_{-2.41}^{+5.43}$ & $1.68_{-0.24}^{+0.60}$ \\
090328A & Band+cutoff & $-1.09_{-0.04}^{+0.08}$ & $-2.00_{-0.10}^{+0.14}$ & $662_{-195}^{+148}$ & $1.53_{-0.37}^{+0.25}$ & $\cdots$ & $46.98_{-15.14}^{+18.36}$ & $\cdots$ & $302.9/394$ & $113_{-36}^{+44}$ & $46.98_{-15.08}^{+18.22}$ & $0.167_{-0.096}^{+0.19}$ \\
090510A & Band+CPL$_{\rm piv}$ & $-0.74_{-0.10}^{+0.12}$ & $-3.48_{-0.78}^{+0.44}$ & $3164_{-639}^{+793}$ & $0.02_{-0.00}^{+0.00}$ & $-1.51_{-0.09}^{+0.59}$ & $2058.67_{-1407.12}^{+3208.13}$ & $0.78_{-0.74}^{+0.49}$ & $337.6/393$ & $1830_{-616}^{+1798}$ & $234.47_{-157.10}^{+987.02}$ & $3.12_{-1.5}^{+8.0}$ \\
090902B & CPL+CPL$_{\rm piv}$ & $-0.55_{-0.04}^{+0.02}$ & $\cdots$ & $532_{-20}^{+27}$ & $1.33_{-0.13}^{+0.25}$ & $-1.93_{-0.01}^{+0.01}$ & $17132.30_{-10542.41}^{+252729.16}$ & $2.01_{-0.09}^{+0.07}$ & $561.7/396$ & $2780_{-384}^{+1529}$ & $29.58_{-25.09}^{+26.33}$ & $26.0_{-6.7}^{+36}$ \\
090926A & Band+cutoff & $-0.70_{-0.03}^{+0.01}$ & $-2.26_{-0.01}^{+0.02}$ & $211_{-5}^{+12}$ & $6.04_{-0.17}^{+0.52}$ & $\cdots$ & $3682.02_{-1778.00}^{+7114.39}$ & $\cdots$ & $610.2/394$ & $1860_{-225}^{+434}$ & $50.87_{-24.55}^{+25.57}$ & $26.7_{-6.0}^{+14}$ \\
091031A & Band+cutoff & $-0.94_{-0.10}^{+0.22}$ & $-2.12_{-0.25}^{+0.24}$ & $512_{-217}^{+226}$ & $0.50_{-0.28}^{+0.23}$ & $\cdots$ & $46.81_{-23.64}^{+95.23}$ & $\cdots$ & $387.1/393$ & $130_{-65}^{+262}$ & $46.81_{-23.47}^{+94.51}$ & $0.118_{-0.091}^{+1.0}$ \\
100116A & Band+cutoff & $-1.09_{-0.10}^{+0.11}$ & $-2.53_{-0.35}^{+0.31}$ & $1187_{-495}^{+902}$ & $0.53_{-0.18}^{+0.28}$ & $\cdots$ & $79.37_{-47.65}^{+642.08}$ & $\cdots$ & $304.4/396$ & $220_{-129}^{+1764}$ & $79.37_{-46.60}^{+637.55}$ & $2.04_{-1.7}^{+166}$ \\
100225A & Band+cutoff & $-0.69_{-0.17}^{+0.26}$ & $-1.99_{-0.19}^{+0.17}$ & $297_{-119}^{+167}$ & $0.20_{-0.12}^{+0.17}$ & $\cdots$ & $62.26_{-29.53}^{+100.45}$ & $\cdots$ & $334.1/395$ & $172_{-81}^{+276}$ & $62.26_{-29.37}^{+99.69}$ & $1.92_{-1.4}^{+12}$ \\
100724B & Band+cutoff & $-0.76_{-0.01}^{+0.04}$ & $-1.97_{-0.04}^{+0.04}$ & $298_{-34}^{+7}$ & $1.01_{-0.15}^{+0.05}$ & $\cdots$ & $33.84_{-3.15}^{+4.16}$ & $\cdots$ & $357.1/397$ & $93.7_{-8.7}^{+11.4}$ & $33.84_{-3.14}^{+4.13}$ & $0.305_{-0.054}^{+0.080}$ \\
100826A & Band+cutoff & $-0.76_{-0.04}^{+0.03}$ & $-1.88_{-0.07}^{+0.06}$ & $213_{-19}^{+16}$ & $1.58_{-0.16}^{+0.23}$ & $\cdots$ & $22.15_{-3.84}^{+4.60}$ & $\cdots$ & $516.4/363$ & $61.3_{-10.6}^{+12.6}$ & $22.15_{-3.83}^{+4.56}$ & $0.103_{-0.033}^{+0.047}$ \\
101123A & Band+cutoff & $-0.89_{-0.03}^{+0.06}$ & $-1.60_{-0.25}^{+0.12}$ & $323_{-74}^{+50}$ & $1.15_{-0.20}^{+0.16}$ & $\cdots$ & $4.24_{-1.76}^{+4.48}$ & $\cdots$ & $326.0/332$ & $11.7_{-4.8}^{+12.3}$ & $4.24_{-1.75}^{+4.44}$ & $(3.81_{-2.5}^{+12})\times10^{-3}$ \\
110328B & Band+cutoff & $-1.20_{-0.10}^{+0.16}$ & $-1.75_{-0.25}^{+0.19}$ & $418_{-132}^{+438}$ & $1.03_{-0.48}^{+0.73}$ & $\cdots$ & $21.45_{-8.40}^{+18.65}$ & $\cdots$ & $362.4/364$ & $59.4_{-23.1}^{+51.2}$ & $21.45_{-8.36}^{+18.51}$ & $0.265_{-0.20}^{+1.2}$ \\
110529A & CPL+CPL$_{\rm piv}$ & $-0.40_{-0.44}^{+0.68}$ & $\cdots$ & $384_{-175}^{+437}$ & $0.19_{-0.18}^{+1.72}$ & $-1.22_{-0.17}^{+2.14}$ & $9.86_{-3.93}^{+8.62}$ & $7.47_{-6.04}^{+5.08}$ & $326.4/363$ & $27.3_{-10.9}^{+23.9}$ & $9.86_{-3.94}^{+8.63}$ & $(1.12_{-0.84}^{+4.8})\times10^{-4}$ \\
110721A & Band+cutoff & $-1.08_{-0.04}^{+0.04}$ & $-2.34_{-0.11}^{+0.08}$ & $815_{-151}^{+162}$ & $3.28_{-0.50}^{+0.49}$ & $\cdots$ & $153.21_{-52.63}^{+251.11}$ & $\cdots$ & $403.7/395$ & $251_{-31}^{+60}$ & $112.78_{-52.19}^{+40.66}$ & $2.63_{-0.61}^{+1.4}$ \\
110731A & CPL+CPL$_{\rm piv}$ & $-0.59_{-0.18}^{+0.20}$ & $\cdots$ & $211_{-37}^{+47}$ & $0.95_{-0.54}^{+1.18}$ & $-1.78_{-0.05}^{+0.11}$ & $218.53_{-67.34}^{+108.66}$ & $0.81_{-0.29}^{+0.18}$ & $404.0/397$ & $1004_{-105}^{+148}$ & $164.11_{-48.35}^{+73.06}$ & $8.85_{-1.7}^{+2.8}$ \\
120226A & Band+cutoff & $-0.95_{-0.03}^{+0.06}$ & $-2.23_{-0.20}^{+0.17}$ & $266_{-44}^{+28}$ & $1.85_{-0.31}^{+0.23}$ & $\cdots$ & $37.30_{-16.82}^{+40.11}$ & $\cdots$ & $303.4/396$ & $103_{-46}^{+110}$ & $37.30_{-16.74}^{+39.80}$ & $0.446_{-0.31}^{+1.5}$ \\
120328B & Band+cutoff & $-0.77_{-0.02}^{+0.05}$ & $-2.03_{-0.05}^{+0.08}$ & $143_{-14}^{+7}$ & $2.91_{-0.39}^{+0.18}$ & $\cdots$ & $20.49_{-9.31}^{+6.71}$ & $\cdots$ & $362.1/391$ & $56.7_{-25.6}^{+18.4}$ & $20.49_{-9.26}^{+6.66}$ & $0.132_{-0.092}^{+0.100}$ \\
120709A & CPL+CPL$_{\rm piv}$ & $-0.78_{-0.35}^{+0.18}$ & $\cdots$ & $313_{-112}^{+172}$ & $0.27_{-0.20}^{+1.42}$ & $-1.94_{-0.14}^{+0.21}$ & $117.19_{-74.35}^{+532.53}$ & $0.12_{-0.09}^{+0.08}$ & $373.6/393$ & $324_{-201}^{+1478}$ & $117.19_{-72.63}^{+533.97}$ & $0.678_{-0.59}^{+22}$ \\
121011A & CPL$_{\rm piv}$ & $\cdots$ & $\cdots$ & $\cdots$ & $\cdots$ & $-1.28_{-0.10}^{+0.07}$ & $9.84_{-3.54}^{+3.64}$ & $0.74_{-0.17}^{+0.14}$ & $279.1/370$ & $21.5_{-7.7}^{+8.0}$ & $9.84_{-3.54}^{+3.65}$ & $(3.88_{-2.8}^{+9.0})\times10^{-4}$ \\
130305A & Band+cutoff & $-0.59_{-0.04}^{+0.06}$ & $-2.36_{-0.29}^{+0.26}$ & $429_{-60}^{+40}$ & $0.34_{-0.07}^{+0.07}$ & $\cdots$ & $36.13_{-22.21}^{+114.64}$ & $\cdots$ & $319.9/365$ & $100.0_{-59.9}^{+315.1}$ & $36.13_{-21.63}^{+113.85}$ & $0.404_{-0.34}^{+6.6}$ \\
130310A & CPL+CPL$_{\rm piv}$ & $0.25_{-0.79}^{+0.74}$ & $\cdots$ & $382_{-208}^{+576}$ & $0.00_{-0.00}^{+0.03}$ & $-1.41_{-0.12}^{+0.10}$ & $12.00_{-5.51}^{+6.39}$ & $2.08_{-0.86}^{+1.08}$ & $371.1/362$ & $33.2_{-15.3}^{+17.7}$ & $12.00_{-5.52}^{+6.39}$ & $(2.11_{-1.5}^{+3.3})\times10^{-4}$ \\
130504C & Band+cutoff & $-1.14_{-0.02}^{+0.04}$ & $-2.07_{-0.35}^{+0.20}$ & $620_{-127}^{+83}$ & $4.67_{-0.67}^{+0.44}$ & $\cdots$ & $15.63_{-5.81}^{+21.00}$ & $\cdots$ & $333.1/398$ & $43.3_{-16.0}^{+57.7}$ & $15.63_{-5.78}^{+20.84}$ & $0.0335_{-0.020}^{+0.15}$ \\
130518A & Band+cutoff & $-0.85_{-0.02}^{+0.04}$ & $-2.50_{-0.12}^{+0.17}$ & $369_{-51}^{+15}$ & $1.53_{-0.20}^{+0.12}$ & $\cdots$ & $73.84_{-38.58}^{+87.51}$ & $\cdots$ & $426.8/396$ & $357_{-184}^{+419}$ & $73.84_{-38.20}^{+86.86}$ & $2.92_{-2.2}^{+11}$ \\
130606B & Band+cutoff & $-1.02_{-0.03}^{+0.02}$ & $-1.80_{-0.07}^{+0.03}$ & $293_{-11}^{+43}$ & $8.20_{-0.62}^{+0.74}$ & $\cdots$ & $7.91_{-1.61}^{+3.89}$ & $\cdots$ & $647.0/333$ & $21.9_{-4.4}^{+10.7}$ & $7.91_{-1.61}^{+3.86}$ & $0.0117_{-0.0043}^{+0.014}$ \\
130828A & CPL+CPL$_{\rm piv}$ & $0.05_{-0.41}^{+0.35}$ & $\cdots$ & $110_{-26}^{+29}$ & $0.01_{-0.01}^{+0.03}$ & $-2.04_{-0.04}^{+0.04}$ & $592.36_{-331.39}^{+1211.51}$ & $0.11_{-0.02}^{+0.01}$ & $467.6/405$ & $472_{-86}^{+125}$ & $49.11_{-27.20}^{+39.83}$ & $6.85_{-1.9}^{+3.5}$ \\
131108A & CPL+CPL$_{\rm piv}$ & $-0.71_{-0.12}^{+0.14}$ & $\cdots$ & $269_{-50}^{+47}$ & $0.83_{-0.38}^{+0.64}$ & $-1.71_{-0.04}^{+0.08}$ & $354.06_{-78.51}^{+107.71}$ & $0.59_{-0.17}^{+0.11}$ & $369.6/397$ & $1666_{-370}^{+507}$ & $354.06_{-78.62}^{+107.84}$ & $4.12_{-1.7}^{+3.2}$ \\
131216A & Band+cutoff & $-0.63_{-0.18}^{+0.31}$ & $-1.89_{-0.40}^{+0.27}$ & $230_{-106}^{+111}$ & $0.17_{-0.09}^{+0.16}$ & $\cdots$ & $15.64_{-10.27}^{+40.27}$ & $\cdots$ & $308.2/365$ & $43.3_{-27.3}^{+110.8}$ & $15.64_{-9.88}^{+40.02}$ & $0.0606_{-0.052}^{+0.72}$ \\
131231A & Band+cutoff & $-1.29_{-0.01}^{+0.03}$ & $-2.42_{-0.12}^{+0.07}$ & $333_{-33}^{+12}$ & $36.53_{-2.99}^{+1.03}$ & $\cdots$ & $37.52_{-13.49}^{+22.44}$ & $\cdots$ & $2964.7/398$ & $85.2_{-30.5}^{+50.6}$ & $37.52_{-13.43}^{+22.27}$ & $0.348_{-0.20}^{+0.54}$ \\
140110A & CPL+CPL$_{\rm piv}$ & $-0.71_{-0.16}^{+0.34}$ & $\cdots$ & $1111_{-591}^{+980}$ & $0.13_{-0.10}^{+0.14}$ & $-1.35_{-0.26}^{+0.94}$ & $143.42_{-91.04}^{+185.50}$ & $0.07_{-0.07}^{+0.12}$ & $379.2/393$ & $384_{-145}^{+237}$ & $133.99_{-93.07}^{+329.21}$ & $5.88_{-3.6}^{+9.4}$ \\
140206B & Band+cutoff & $-1.30_{-0.00}^{+0.00}$ & $-1.98_{-0.08}^{+0.07}$ & $475_{-33}^{+19}$ & $5.61_{-0.10}^{+0.10}$ & $\cdots$ & $39.73_{-3.14}^{+4.26}$ & $\cdots$ & $393.6/396$ & $110_{-9}^{+12}$ & $39.73_{-3.13}^{+4.23}$ & $0.647_{-0.10}^{+0.15}$ \\
140619B & CPL+CPL$_{\rm piv}$ & $-0.29_{-0.42}^{+0.48}$ & $\cdots$ & $888_{-483}^{+897}$ & $0.01_{-0.01}^{+0.07}$ & $-1.46_{-0.16}^{+0.63}$ & $325.63_{-204.20}^{+568.51}$ & $0.17_{-0.16}^{+0.16}$ & $421.5/393$ & $901_{-553}^{+1577}$ & $325.63_{-199.92}^{+569.97}$ & $2.07_{-1.9}^{+892}$ \\
141028A & Band+cutoff & $-0.87_{-0.04}^{+0.07}$ & $-2.02_{-0.07}^{+0.13}$ & $290_{-59}^{+38}$ & $1.24_{-0.26}^{+0.22}$ & $\cdots$ & $48.08_{-17.33}^{+16.33}$ & $\cdots$ & $315.9/395$ & $222_{-80}^{+75}$ & $48.08_{-17.26}^{+16.21}$ & $1.51_{-0.89}^{+1.2}$ \\
141207A & CPL+CPL$_{\rm piv}$ & $-0.03_{-0.14}^{+0.15}$ & $\cdots$ & $377_{-55}^{+59}$ & $0.01_{-0.01}^{+0.01}$ & $-1.81_{-0.03}^{+0.03}$ & $1998.16_{-1179.39}^{+9191.02}$ & $0.32_{-0.04}^{+0.04}$ & $367.8/395$ & $530_{-70}^{+157}$ & $18.39_{-13.04}^{+16.22}$ & $11.1_{-2.6}^{+7.3}$ \\
141222A & CPL+CPL$_{\rm piv}$ & $-1.26_{-0.18}^{+0.36}$ & $\cdots$ & $2238_{-1457}^{+2781}$ & $8.38_{-7.33}^{+17.23}$ & $-1.83_{-0.14}^{+0.12}$ & $142.78_{-91.35}^{+10245.17}$ & $2.00_{-1.22}^{+2.30}$ & $385.4/395$ & $395_{-247}^{+28432}$ & $142.78_{-89.07}^{+10273.49}$ & $0.0619_{-0.053}^{+332}$ \\
150210A & CPL+CPL$_{\rm piv}$ & $-0.83_{-0.08}^{+0.10}$ & $\cdots$ & $2084_{-379}^{+368}$ & $0.27_{-0.12}^{+0.16}$ & $-1.82_{-0.12}^{+0.12}$ & $115.13_{-73.77}^{+434.78}$ & $0.20_{-0.09}^{+0.10}$ & $337.3/391$ & $319_{-199}^{+1207}$ & $115.13_{-71.89}^{+435.99}$ & $0.413_{-0.36}^{+9.2}$ \\
150416A & CPL+CPL$_{\rm piv}$ & $-0.69_{-0.16}^{+0.21}$ & $\cdots$ & $448_{-148}^{+227}$ & $0.12_{-0.03}^{+0.07}$ & $-1.18_{-0.37}^{+0.60}$ & $72.83_{-47.80}^{+129.22}$ & $0.03_{-0.03}^{+2.50}$ & $370.4/366$ & $202_{-128}^{+359}$ & $72.83_{-46.39}^{+129.60}$ & $4.81_{-4.2}^{+33}$ \\
150523A & CPL+CPL$_{\rm piv}$ & $-0.26_{-0.24}^{+0.23}$ & $\cdots$ & $262_{-57}^{+83}$ & $0.02_{-0.01}^{+0.03}$ & $-1.97_{-0.05}^{+0.06}$ & $657.67_{-487.02}^{+9642.61}$ & $0.09_{-0.02}^{+0.03}$ & $298.5/397$ & $362_{-70}^{+218}$ & $25.98_{-21.98}^{+33.33}$ & $7.32_{-2.4}^{+11}$ \\
151006A & CPL+CPL$_{\rm piv}$ & $-0.34_{-0.95}^{+3.09}$ & $\cdots$ & $73_{-54}^{+264}$ & $0.02_{-0.02}^{+1.23}$ & $-1.44_{-0.11}^{+1.08}$ & $7.77_{-3.71}^{+2.47}$ & $0.76_{-0.34}^{+0.20}$ & $275.7/375$ & $21.5_{-10.3}^{+6.8}$ & $7.77_{-3.71}^{+2.47}$ & $0.0679_{-0.050}^{+0.057}$ \\
160509A & Band+cutoff & $-0.79_{-0.09}^{+0.05}$ & $-2.20_{-0.04}^{+0.04}$ & $326_{-26}^{+94}$ & $0.36_{-0.05}^{+0.14}$ & $\cdots$ & $96.72_{-16.46}^{+18.02}$ & $\cdots$ & $396.9/407$ & $290_{-49}^{+54}$ & $96.72_{-16.39}^{+17.89}$ & $1.82_{-0.61}^{+0.87}$ \\
160625B & CPL+CPL$_{\rm piv}$ & $-0.60_{-0.06}^{+0.14}$ & $\cdots$ & $418_{-81}^{+43}$ & $0.21_{-0.09}^{+0.09}$ & $-1.79_{-0.02}^{+0.01}$ & $75.23_{-5.37}^{+10.13}$ & $0.37_{-0.02}^{+0.02}$ & $664.6/404$ & $250_{-18}^{+34}$ & $75.23_{-5.37}^{+10.15}$ & $0.854_{-0.13}^{+0.25}$ \\
160709A & CPL+CPL$_{\rm piv}$ & $-0.49_{-0.24}^{+0.27}$ & $\cdots$ & $1522_{-478}^{+688}$ & $0.03_{-0.02}^{+0.09}$ & $-1.65_{-0.06}^{+0.09}$ & $226.29_{-87.89}^{+203.52}$ & $0.40_{-0.12}^{+0.11}$ & $360.3/403$ & $626_{-244}^{+564}$ & $226.29_{-88.01}^{+203.77}$ & $0.308_{-0.25}^{+2.4}$ \\
160816A & CPL+CPL$_{\rm piv}$ & $-0.69_{-0.03}^{+0.04}$ & $\cdots$ & $181_{-12}^{+9}$ & $2.29_{-0.32}^{+0.24}$ & $-1.04_{-0.67}^{+0.75}$ & $84.83_{-62.54}^{+155.75}$ & $0.01_{-0.01}^{+0.13}$ & $391.5/404$ & $235_{-162}^{+433}$ & $84.83_{-58.63}^{+156.52}$ & $0.118_{-0.11}^{+0.85}$ \\
160821A & Band+cutoff & $-0.97_{-0.00}^{+0.00}$ & $-2.18_{-0.01}^{+0.00}$ & $917_{-19}^{+6}$ & $10.32_{-0.05}^{+0.03}$ & $\cdots$ & $29.02_{-1.12}^{+0.49}$ & $\cdots$ & $686.6/403$ & $80.3_{-3.1}^{+1.3}$ & $29.02_{-1.12}^{+0.48}$ & $0.210_{-0.038}^{+0.041}$ \\
160905A & CPL+CPL$_{\rm piv}$ & $-0.77_{-0.10}^{+0.14}$ & $\cdots$ & $729_{-193}^{+116}$ & $0.72_{-0.35}^{+0.57}$ & $-1.43_{-0.21}^{+1.41}$ & $11.01_{-4.36}^{+10.76}$ & $0.85_{-0.44}^{+1.11}$ & $418.0/405$ & $30.5_{-12.1}^{+29.8}$ & $11.01_{-4.36}^{+10.77}$ & $0.0383_{-0.024}^{+0.11}$ \\
160910A & Band+cutoff & $-0.74_{-0.03}^{+0.04}$ & $-2.11_{-0.06}^{+0.07}$ & $251_{-27}^{+19}$ & $1.99_{-0.25}^{+0.24}$ & $\cdots$ & $48.62_{-13.36}^{+20.49}$ & $\cdots$ & $470.2/376$ & $135_{-37}^{+56}$ & $48.62_{-13.30}^{+20.33}$ & $0.362_{-0.17}^{+0.37}$ \\
170214A & Band+cutoff & $-0.86_{-0.02}^{+0.03}$ & $-2.41_{-0.03}^{+0.01}$ & $348_{-30}^{+25}$ & $1.35_{-0.05}^{+0.06}$ & $\cdots$ & $339.81_{-71.04}^{+166.79}$ & $\cdots$ & $410.5/407$ & $1121_{-72}^{+110}$ & $154.86_{-34.57}^{+27.77}$ & $11.9_{-1.4}^{+2.4}$ \\
170405A & Band+cutoff & $-0.31_{-0.10}^{+0.09}$ & $-2.14_{-0.10}^{+0.05}$ & $122_{-13}^{+16}$ & $0.14_{-0.04}^{+0.06}$ & $\cdots$ & $48.29_{-9.47}^{+29.23}$ & $\cdots$ & $327.9/403$ & $301_{-59}^{+181}$ & $48.29_{-9.43}^{+29.00}$ & $1.53_{-0.56}^{+2.4}$ \\
170424A & Band+cutoff & $-0.98_{-0.08}^{+0.10}$ & $-2.05_{-0.21}^{+0.22}$ & $577_{-172}^{+202}$ & $0.70_{-0.22}^{+0.25}$ & $\cdots$ & $27.62_{-12.22}^{+29.48}$ & $\cdots$ & $327.5/406$ & $76.4_{-33.7}^{+81.0}$ & $27.62_{-12.16}^{+29.26}$ & $0.0662_{-0.055}^{+0.79}$ \\
180703A & Band+cutoff & $-0.83_{-0.07}^{+0.11}$ & $-1.83_{-0.17}^{+0.14}$ & $311_{-92}^{+98}$ & $0.69_{-0.21}^{+0.19}$ & $\cdots$ & $23.64_{-9.64}^{+18.62}$ & $\cdots$ & $319.6/404$ & $54.6_{-22.2}^{+42.6}$ & $23.64_{-9.60}^{+18.48}$ & $0.216_{-0.14}^{+0.47}$ \\
180720B & Band+cutoff & $-1.11_{-0.00}^{+0.00}$ & $-2.18_{-0.23}^{+0.02}$ & $656_{-66}^{+5}$ & $12.89_{-0.07}^{+0.06}$ & $\cdots$ & $18.36_{-1.89}^{+16.45}$ & $\cdots$ & $706.8/403$ & $42.0_{-4.3}^{+37.4}$ & $18.36_{-1.88}^{+16.33}$ & $(7.88_{-2.4}^{+22})\times10^{-3}$ \\
190114C & CPL+CPL$_{\rm piv}$ & $-0.61_{-0.04}^{+0.03}$ & $\cdots$ & $528_{-35}^{+7}$ & $0.47_{-0.06}^{+0.12}$ & $-1.90_{-0.01}^{+0.01}$ & $9467.55_{-6704.17}^{+29896.88}$ & $1.10_{-0.03}^{+0.05}$ & $376.3/403$ & $1423_{-225}^{+356}$ & $55.37_{-34.66}^{+64.74}$ & $12.5_{-3.6}^{+7.0}$ \\
190122A & CPL$_{\rm piv}$ & $\cdots$ & $\cdots$ & $\cdots$ & $\cdots$ & $-1.56_{-0.04}^{+0.05}$ & $18.42_{-3.32}^{+3.68}$ & $0.79_{-0.10}^{+0.12}$ & $280.8/377$ & $51.0_{-9.2}^{+10.2}$ & $18.42_{-3.33}^{+3.68}$ & $0.0841_{-0.029}^{+0.042}$ \\
190531B & CPL+CPL$_{\rm piv}$ & $-0.65_{-0.06}^{+0.00}$ & $\cdots$ & $389_{-58}^{+29}$ & $0.51_{-0.01}^{+0.15}$ & $-1.69_{-0.07}^{+0.02}$ & $13.49_{-3.06}^{+5.91}$ & $0.65_{-0.05}^{+0.07}$ & $1869.2/406$ & $37.3_{-8.5}^{+16.4}$ & $13.49_{-3.06}^{+5.92}$ & $0.0233_{-0.0094}^{+0.025}$ \\
200101A & Band+cutoff & $-0.75_{-0.02}^{+0.02}$ & $-2.21_{-0.08}^{+0.09}$ & $472_{-29}^{+26}$ & $3.38_{-0.25}^{+0.26}$ & $\cdots$ & $32.68_{-14.35}^{+18.46}$ & $\cdots$ & $417.6/365$ & $90.4_{-39.5}^{+50.7}$ & $32.68_{-14.29}^{+18.32}$ & $0.0855_{-0.059}^{+0.12}$ \\
200313A & Band+cutoff & $-0.56_{-0.04}^{+0.05}$ & $-1.75_{-0.06}^{+0.05}$ & $178_{-19}^{+18}$ & $1.49_{-0.21}^{+0.22}$ & $\cdots$ & $13.53_{-4.17}^{+8.26}$ & $\cdots$ & $403.6/332$ & $37.5_{-11.5}^{+22.7}$ & $13.53_{-4.15}^{+8.19}$ & $0.0342_{-0.018}^{+0.054}$ \\
200412B & Band+cutoff & $-0.39_{-0.03}^{+0.03}$ & $-2.36_{-0.05}^{+0.05}$ & $166_{-8}^{+8}$ & $1.90_{-0.18}^{+0.18}$ & $\cdots$ & $101.71_{-36.65}^{+67.85}$ & $\cdots$ & $496.5/402$ & $281_{-101}^{+186}$ & $101.71_{-36.50}^{+67.33}$ & $0.604_{-0.36}^{+1.1}$ \\
200829A & Band+cutoff & $-0.38_{-0.03}^{+0.04}$ & $-2.27_{-0.07}^{+0.12}$ & $212_{-14}^{+9}$ & $3.36_{-0.40}^{+0.39}$ & $\cdots$ & $10.63_{-4.75}^{+8.43}$ & $\cdots$ & $781.6/335$ & $33.1_{-14.7}^{+26.1}$ & $10.63_{-4.73}^{+8.37}$ & $0.0214_{-0.015}^{+0.047}$ \\
201104A & CPL+CPL$_{\rm piv}$ & $-0.91_{-0.15}^{+0.59}$ & $\cdots$ & $1053_{-657}^{+928}$ & $0.24_{-0.23}^{+0.23}$ & $-1.45_{-0.28}^{+0.61}$ & $292.67_{-171.12}^{+486.28}$ & $0.04_{-0.04}^{+0.11}$ & $326.1/402$ & $810_{-469}^{+1348}$ & $292.67_{-169.31}^{+487.19}$ & $6.54_{-5.4}^{+41}$ \\
210410A & CPL+CPL$_{\rm piv}$ & $-0.54_{-0.21}^{+0.15}$ & $\cdots$ & $739_{-93}^{+326}$ & $0.07_{-0.04}^{+0.13}$ & $-1.53_{-0.28}^{+0.17}$ & $50.31_{-16.60}^{+567.99}$ & $0.28_{-0.22}^{+0.22}$ & $230.1/362$ & $139_{-46}^{+1574}$ & $50.31_{-16.63}^{+568.67}$ & $0.259_{-0.15}^{+40}$ \\
210619B & Band+cutoff & $-0.79_{-0.01}^{+0.01}$ & $-1.86_{-0.06}^{+0.06}$ & $159_{-7}^{+6}$ & $5.92_{-0.28}^{+0.26}$ & $\cdots$ & $3.73_{-0.89}^{+1.23}$ & $\cdots$ & $432.4/333$ & $15.2_{-3.6}^{+4.9}$ & $3.73_{-0.88}^{+1.22}$ & $(2.03_{-0.85}^{+1.6})\times10^{-3}$ \\
220101A & Band+cutoff & $-0.95_{-0.10}^{+0.10}$ & $-2.35_{-0.08}^{+0.10}$ & $276_{-59}^{+84}$ & $0.98_{-0.31}^{+0.39}$ & $\cdots$ & $141.87_{-63.85}^{+110.71}$ & $\cdots$ & $358.8/402$ & $1103_{-494}^{+854}$ & $141.87_{-63.54}^{+109.87}$ & $3.87_{-3.1}^{+19}$ \\
220107B & CPL+CPL$_{\rm piv}$ & $-0.78_{-0.15}^{+0.10}$ & $\cdots$ & $319_{-83}^{+79}$ & $0.61_{-0.35}^{+0.62}$ & $-1.81_{-0.09}^{+0.56}$ & $431.60_{-257.47}^{+1183.17}$ & $0.14_{-0.13}^{+0.07}$ & $351.1/403$ & $529_{-126}^{+460}$ & $84.74_{-55.29}^{+247.29}$ & $4.36_{-1.8}^{+11}$ \\
220209A & Band+cutoff & $-0.95_{-0.10}^{+0.11}$ & $-2.28_{-0.12}^{+0.10}$ & $417_{-119}^{+144}$ & $0.42_{-0.13}^{+0.21}$ & $\cdots$ & $138.92_{-55.87}^{+156.27}$ & $\cdots$ & $342.3/408$ & $349_{-63}^{+94}$ & $114.16_{-50.65}^{+71.49}$ & $7.38_{-2.3}^{+4.4}$ \\
220527A & Band+CPL$_{\rm piv}$ & $-0.65_{-0.06}^{+0.13}$ & $-2.75_{-0.15}^{+0.10}$ & $114_{-13}^{+9}$ & $0.25_{-0.01}^{+0.02}$ & $-1.54_{-0.16}^{+0.47}$ & $37.64_{-11.29}^{+14.82}$ & $1.02_{-0.78}^{+0.67}$ & $400.2/404$ & $96.7_{-29.0}^{+38.1}$ & $37.64_{-11.30}^{+14.84}$ & $0.0866_{-0.044}^{+0.082}$ \\
221023A & CPL+CPL$_{\rm piv}$ & $-1.03_{-0.01}^{+0.01}$ & $\cdots$ & $976_{-64}^{+30}$ & $9.44_{-0.70}^{+0.44}$ & $-1.29_{-0.07}^{+0.04}$ & $36.88_{-1.52}^{+5.70}$ & $2.09_{-0.28}^{+0.32}$ & $679.4/364$ & $102_{-4}^{+16}$ & $36.88_{-1.52}^{+5.70}$ & $0.251_{-0.033}^{+0.091}$ \\
221027A & Band+cutoff & $-0.59_{-0.12}^{+0.08}$ & $-2.22_{-0.19}^{+0.16}$ & $219_{-48}^{+58}$ & $0.37_{-0.16}^{+0.20}$ & $\cdots$ & $72.25_{-40.36}^{+242.54}$ & $\cdots$ & $350.8/406$ & $200_{-110}^{+666}$ & $72.25_{-39.80}^{+240.75}$ & $0.711_{-0.57}^{+13}$ \\
221209A & Band+cutoff & $-0.54_{-0.06}^{+0.06}$ & $-1.87_{-0.13}^{+0.11}$ & $196_{-24}^{+27}$ & $1.94_{-0.35}^{+0.41}$ & $\cdots$ & $6.67_{-2.62}^{+7.10}$ & $\cdots$ & $429.7/336$ & $18.5_{-7.2}^{+19.5}$ & $6.67_{-2.61}^{+7.04}$ & $(4.07_{-2.6}^{+13})\times10^{-3}$ \\
230512A & CPL+CPL$_{\rm piv}$ & $-0.63_{-0.05}^{+0.06}$ & $\cdots$ & $973_{-126}^{+138}$ & $0.56_{-0.15}^{+0.17}$ & $-1.22_{-0.28}^{+1.38}$ & $67.60_{-50.53}^{+92.01}$ & $0.18_{-0.18}^{+0.37}$ & $405.0/404$ & $187_{-131}^{+256}$ & $67.60_{-47.16}^{+92.50}$ & $0.0772_{-0.071}^{+0.45}$ \\
240118A & Band+cutoff & $-0.70_{-0.00}^{+0.00}$ & $-2.48_{-0.03}^{+0.02}$ & $305_{-4}^{+5}$ & $4.18_{-0.01}^{+0.01}$ & $\cdots$ & $56.77_{-7.32}^{+13.56}$ & $\cdots$ & $727.5/405$ & $132_{-17}^{+31}$ & $56.77_{-7.29}^{+13.46}$ & $0.193_{-0.047}^{+0.10}$ \\
240825A & Band+CPL$_{\rm piv}$ & $-0.82_{-0.02}^{+0.03}$ & $-2.40_{-0.06}^{+0.06}$ & $397_{-19}^{+17}$ & $0.35_{-0.01}^{+0.01}$ & $-1.20_{-0.17}^{+0.36}$ & $72.54_{-25.76}^{+27.49}$ & $2.57_{-1.62}^{+1.81}$ & $421.2/360$ & $167_{-59}^{+63}$ & $72.54_{-25.80}^{+27.52}$ & $0.0815_{-0.048}^{+0.076}$ \\
240905B & CPL+CPL$_{\rm piv}$ & $0.18_{-0.21}^{+0.15}$ & $\cdots$ & $446_{-71}^{+93}$ & $0.01_{-0.01}^{+0.02}$ & $-1.48_{-0.33}^{+0.12}$ & $39.14_{-16.03}^{+174.64}$ & $2.57_{-1.94}^{+1.66}$ & $371.6/363$ & $108_{-44}^{+484}$ & $39.14_{-16.05}^{+174.85}$ & $0.0285_{-0.019}^{+0.83}$ \\
250716A & Band+cutoff & $-1.18_{-0.07}^{+0.09}$ & $-1.58_{-0.15}^{+0.09}$ & $546_{-239}^{+389}$ & $4.30_{-1.08}^{+1.01}$ & $\cdots$ & $11.28_{-4.11}^{+12.54}$ & $\cdots$ & $307.4/333$ & $21.5_{-7.8}^{+23.7}$ & $11.28_{-4.09}^{+12.45}$ & $0.0397_{-0.024}^{+0.14}$ \\
\end{longtable}
\noindent\footnotesize Only the model with the lowest BIC among the acceptable fits is listed.
The quoted uncertainties are 90\% confidence intervals,
and ellipses denote parameters not applicable to the adopted model.
$N_0$ is the normalization of the Band or low-energy CPL component,
whereas $N_1$ is the CPL$_{\rm piv}$ normalization at 1~MeV in units of
$10^{-4}~{\rm photons~cm^{-2}~s^{-1}~keV^{-1}}$.
\par\endgroup
\end{landscape}

%% file: jcapexample.bbl
\providecommand{\href}[2]{#2}\begingroup\raggedright\endgroup